\documentclass[11pt,xcolor=dvipsnames]{article}
\DeclareMathAlphabet{\scr}{U}{rsfs}{m}{n}

\pdfoutput=1
\usepackage{scalerel,stackengine}
\usepackage[colorlinks,urlcolor=black,linkcolor = blue,citecolor = black,]{hyperref}

\usepackage{subfigure}

\usepackage{latexsym}
\usepackage{epsfig}
\usepackage[mathscr]{eucal}
\usepackage{amsfonts}
\usepackage{amscd}
\usepackage{cite}
\usepackage{array}
\usepackage{amssymb}
\usepackage{colordvi}
\usepackage[centertags]{amsmath}
\usepackage{enumerate}
\usepackage{graphicx}
\usepackage{booktabs}
\usepackage{theorem}
\usepackage[footnotesize]{caption}
\usepackage{soul}
\usepackage{mcite}
\usepackage{slashed}%slash notation for gamma matrices
\usepackage{braket}%for bra and ket notations

\usepackage{xcolor}
\usepackage{ulem}
\usepackage{bbm}
\usepackage[utf8]{inputenc}
\usepackage{fancyvrb}% provides improved Verbatim command
\usepackage{framed}
\usepackage{xspace}
\usepackage{todonotes}
\newcommand{\newc}{\newcommand}
\newc{\be}{\begin{equation}}
\newc{\bea}{\begin{eqnarray}}
\newc{\eea}{\end{eqnarray}}
\newc{\ol}{\overline}
\newc{\wt}{\widetilde}
\newc{\bs}{\boldsymbol}
\newc{\m}{\mathcal}
\newc{\la}{\langle}
\newc{\ra}{\rangle}

\newcommand{\beq}{\begin{eqnarray}}
\newcommand{\eeq}{\end{eqnarray}}
\newcommand{\bpmatrix}{\begin{pmatrix}}
\newcommand{\epmatrix}{\end{pmatrix}}
\newcommand{\ba}{\begin{array}}
\newcommand{\ea}{\end{array}}

\renewcommand{\ol}{\text{1l}}

\renewcommand{\Re}{\text{Re}\!}

\renewcommand{\eqref}[1]{Eq.~(\ref{#1})}

\newcommand{\bc}{\begin{center}}
\newcommand{\ec}{\end{center}}

\newcommand{\s}{\newline \vspace*{-3.5mm}}

\newcommand{\nZ}{\mathbb{Z}}

\newcommand{\ii}{\mathit{i}}

\newcommand{\figref}[2][{}]{\hyperref[#2]{\figurename~\ref{#2}#1}}

\newenvironment{kasten*}[1]
{
\hspace{0.05\linewidth}
\begin{minipage}{0.95\linewidth}
\setlength{\fboxsep}{10pt}
\definecolor{shadecolor}{gray}{0.9}
\definecolor{framecolor}{gray}{0}

\MakeFramed {\FrameRestore}
\subsection*{#1}
}
{
\endMakeFramed
\end{minipage}
\vspace{1em}
}

\newcommand{\cbrak}[1]{\left(#1\right)}
\newcommand{\sbrak}[1]{\left[#1\right]}

\newcommand{\X}{\chi}
\newcommand{\mX}{m_{\chi}}

\newcommand{\gX}{g_{\chi}}
\newcommand{\GX}{G^{\chi}}
\newcommand{\mh}{m_{h_1}}
\newcommand{\mH}{m_{h_2}}

\allowdisplaybreaks

\usepackage[printonlyused]{acronym}
\usepackage{cleveref}
\crefname{chapter}{Chapter}{Chapter}
\crefname{section}{Sec.}{Secs.}
\crefname{table}{Tab.}{Tabs.}
\crefname{figure}{Fig.}{Figs.}
\crefname{equation}{Eq.}{Eqs.}
\crefname{appendix}{Appendix\ }{Appendix\ }

\begin{document}
\pdfoutput=1

%%%%%%%%%%%%%%%%%%%%%%%%%%%%%%%%%%%%%%%%%%%%%%%%%%%%%%%%%%%%%%
%\begin{acronym}
%\acro{VDM}{Vector Dark Matter}
%\acro{DM}{Dark Matter}
%\acro{LO}{Leading Order}
%\acro{NLO}{Next-to-LO}
%\acro{SM}{Standard Model}
%\acro{BSM}{Beyond the SM}
%\acro{EW}{Electroweak}
%\acro{VEV}{Vacuum Expectation Value}
%\acro{2HDM}{Two-Higgs-Doublet-Model}
%\acro{N2HDM}{Next-to-2HDM}
%\acro{OS}{on-shell}
%\acro{IR}{infra-red}
%\acro{QED}{Quantum-electro-dynamics}
%\acro{SI}{spin-independent}
%\acro{SD}{spin-dependent}
%\end{acronym}
%%%%%%%%%%%%%%%%%%%%%%%%%%%%%%%%%%%%%%%%%%%%%%%%%%%%%%%%%%%%%%
\title{
\vspace*{-3.7cm}
\phantom{h} \hfill\mbox{\small KA-TP-16-2019}\\[-1.1cm]
\vspace*{2.7cm}
\textbf{Electroweak Corrections to \\ Dark Matter Direct Detection \\
in a Vector Dark Matter Model \\[4mm]}}

\date{}
\author{
Seraina Glaus$^{1,2\,}$\footnote{E-mail:
\texttt{seraina.glaus@kit.edu}} ,
Margarete M\"{u}hlleitner$^{1\,}$\footnote{E-mail:
\texttt{milada.muehlleitner@kit.edu}} ,
Jonas M\"{u}ller$^{1\,}$\footnote{E-mail:
\texttt{jonas.mueller@kit.edu}} ,
\\
Shruti Patel$^{1,2\,}$\footnote{E-mail: \texttt{shruti.patel@kit.edu}},
Rui Santos$^{3,4\,}$\footnote{E-mail:
  \texttt{rasantos@fc.ul.pt}} 
\\[5mm]
{\small\it
$^1$Institute for Theoretical Physics, Karlsruhe Institute of Technology,} \\
{\small\it 76128 Karlsruhe, Germany}\\[3mm]
{\small\it
$^2$Institute for Nuclear Physics, Karlsruhe Institute of Technology, 
76344 Karlsruhe, Germany}\\[3mm]
{\small\it $^3$Centro de F\'{\i}sica Te\'{o}rica e Computacional,
    Faculdade de Ci\^{e}ncias,} \\
{\small \it    Universidade de Lisboa, Campo Grande, Edif\'{\i}cio C8
  1749-016 Lisboa, Portugal} \\[3mm]
{\small\it
$^4$ISEL -
 Instituto Superior de Engenharia de Lisboa,} \\
{\small \it   Instituto Polit\'ecnico de Lisboa
 1959-007 Lisboa, Portugal} \\[3mm]
}

\maketitle

\begin{abstract}
Although many astrophysical and cosmological observations point
towards the existence of Dark Matter (DM), the nature of the DM particle 
has not been clarified to date. In this paper, we investigate a
minimal model with a vector DM (VDM) candidate. Within this model, we
compute the cross section for the scattering of the VDM particle with
a nucleon. We provide the next-to-leading order (NLO) cross section
for the direct detection of the DM particle. Subsequently, we study
the phenomenological implications of the NLO corrections, in
particular with respect to the sensitivity of the direct detection DM 
experiments. We further investigate more theoretical questions such as the
gauge dependence of the results and the remaining theoretical
uncertainties due to the applied approximations. 
\end{abstract}
\thispagestyle{empty}
\vfill
\newpage
\setcounter{page}{1}

%%%%%%%%%%%%%%%%%%%%%%%%%%%%%%%%%%%%%%%%%%%%%%%%%%%%%%%
\section{Introduction\label{sec:Intro}}
The first indirect hints of the existence of Dark Matter (DM) were reported more than 80 years ago\cite{Zwicky:1933gu} (see \cite{Bertone:2016nfn} for a historical
account). Over the years, confirmations from different sources have
established the existence of DM. Still, as of today, these are all
indirect evidences and all based on gravitational effects. Therefore,
it may come as a surprise that today the discussion on the properties
of DM is about particles which is a consequence of the findings from
astronomy and cosmology. Direct, indirect and collider
searches for DM refer in most cases to particles belonging to some
extension of the Standard Model (SM), and all experimental data from
the different sources favour a weakly interacting massive particle
(WIMP) with a velocity of the order of 200 km/s. That is, DM is
non-relativistic. \s

Many experiments have been proposed for the direct detection of DM on
Earth. It was shown in Ref.~\cite{Goodman:1984dc} that DM particles that
undergo coherent scattering with nuclei, {\it i.e.}~spin-independent
scattering, are the ones with larger scattering rates, and therefore
they can be detected more easily. The scattering rates depend on the material of
the detector, on the underlying cosmological model through the
assumption of an approximately constant DM density, $\rho_0 =
0.3$ GeV/cm$^3$, on the DM velocity and finally on the 
DM-nucleon cross section. Although there are uncertainties associated
with the determination of all the parameters, the need for an increased precision in the DM-nucleon cross section calculation has led several
groups to invest in the calculation of higher-order corrections, both
strong and electroweak, to the scattering cross
section~\cite{Haisch:2013uaa, Crivellin:2014qxa, Hill:2014yka,
  Abe:2015rja, Klasen:2016qyz, Azevedo:2018exj, Ishiwata:2018sdi,
  Ghorbani:2018pjh, Abe:2018emu, Ertas:2019dew}. \s

Although the hypothesis of DM as a particle is now the strongest and
most intensely studied conjecture to explain the data, there are no hints
on the exact nature of the particle itself.
Among the several possibilities, in this study we will focus on a
minimal model with a vector DM candidate. The model is a very simple
extension of the SM where a dark vector $\chi_\mu$ with a gauged
$U(1)_{\chi}$ symmetry and a complex SM-gauge singlet $S$ are added to
the SM field content. We end up with a vector DM (VDM) candidate
$\chi_\mu$ and a new CP-even scalar that mixes with the SM scalar field
coming from the doublet. \s

The electroweak corrections to the coherent scattering of the DM
candidate $\chi_\mu$ first require the renormalisation of the VDM
model and second, the extraction of the spin-independent contributions
from the loop corrections to the effective couplings of the 
Lagrangian, ${\cal L}_{\rm eff}$, which couple two DM
particles and two quarks. These will then constitute the corrections
to the tree-level effective couplings from ${\cal L}_{\rm eff}$. \s

The paper is organised as follows: in Section~\ref{sec:VDM}  we
present the VDM model and in Section~\ref{sec:Renorm} we describe its
renormalisation. In Section~\ref{sec:ddtree} we
discuss the scattering of scalar DM off nuclei at leading order (LO) 
while in Section~\ref{sec:dd1loop} we calculate the electroweak
corrections to the cross section. In Section~\ref{sec:result} we
present and discuss our results. In the conclusions,
Section~\ref{sec:conclusion}, we summarise our findings. Feynman rules
and technical details are left to the appendices. 

%%%%%%%%%%%%%%%%%%%%%%%%%%%%%%%%%%%%%%%%%%%%%%%%%%%%%%%%%%%%%
\section{The Vector Dark Matter Model\label{sec:VDM}}

The VDM model discussed in this work is an extension of the SM, where
a complex SM-gauge singlet $S$ is added to the SM field content
\cite{Hambye:2008bq, Lebedev:2011iq, Farzan:2012hh, Baek:2012se,
  Baek:2014jga, Duch:2015jta, Azevedo:2018oxv}. The model has a new
$U(1)_{\X}$  gauged symmetry, under which solely the gauge singlet $S$
is charged. As the symmetry is gauged, a new vector boson appears in
the theory, which is denoted by $\chi_\mu$. \s 

In order to obtain a stable VDM candidate we assume a $\nZ_2$
symmetry. The dark gauge boson $\chi_{\mu}$ and the scalar field $S$ 
transform under the $\nZ_2$ symmetry as follows 
\begin{equation}
	\chi_{\mu} \rightarrow -\chi_{\mu} \qquad \text{and} \qquad S\rightarrow S^*\,,
\end{equation}
and the SM particles are all even under $\nZ_2$, which precludes
kinetic mixing between the gauge bosons from $U(1)_{\X}$ and the
SM $U(1)_Y$. As the singlet $S$ is charged under the
dark $U(1)_{\X}$, its covariant derivative reads
\begin{equation}
    \mathcal{D}_{\mu}S=\left(\partial_{\mu}+  \mathit{i} \gX \chi_{\mu}\right)S\,,
\end{equation}
where $\gX$ is the gauge coupling of the dark gauge boson $\chi_{\mu}$. \s

The most general Higgs potential invariant under the SM and the
$\nZ_2$ symmetries can be written as  
\begin{equation}
    V= -\mu_H^2 \vert H\vert^2 +\lambda_H \vert
    H\vert^4 - \mu_S^2 \vert 
    S \vert ^2 + \lambda_S \vert S\vert^4 +\kappa \vert S\vert^2 \vert
    H\vert^2\,, 
    \label{VDM::potential}
\end{equation}
in terms of the squared mass parameters $\mu_H^2$,
  $\mu_S^2$ and the quartic couplings $\lambda_H$, $\lambda_S$ and $\kappa$.
The neutral component of the Higgs doublet $H$ and the real part of
the singlet field each acquire a vacuum expectation value (VEV) $v$ and
$v_S$, respectively. The expansions around their VEVs can be written as
\begin{equation}
	H = 
	\begin{pmatrix}
	G^{+}\\\frac{1}{\sqrt 2}\cbrak{v + \Phi_H +\mathit{i} \sigma_H}
	\end{pmatrix} \quad \mbox{and} \quad
	S = \frac{1}{\sqrt{2}} \left( v_S + \Phi_S+ \mathit{i} \sigma_S\right)\,,
\end{equation}
where $\Phi_H$ and $\Phi_S$ denote the CP-even field components of $H$ and $S$.
The CP-odd field components $\sigma_H$ and $\sigma_S$ 
do not acquire VEVs and are therefore identified with the neutral
SM-like Goldstone boson $G^0$ and the Goldstone boson $\GX$ for the
gauge boson $\chi_{\mu}$, respectively, while $G^{\pm}$ are the Goldstone
bosons of the W bosons. The minimum conditions of the
potential yield the tadpole equations
\begin{align}
	\left< \frac{\partial V}{\partial \Phi_H}\right>\equiv
  \frac{T_{\Phi_H}}{v} &= \left(\frac{\kappa v_S^2 }{2}+\lambda_H
                         v^2-\mu_H^2\right)\,, \\
    \left<\frac{\partial V}{\partial \Phi_S}\right>\equiv
  \frac{T_{\Phi_S}}{v_S} &= \left(\frac{\kappa v^2 }{2}+\lambda_S
                           v_S^2-\mu_S^2\right)\,, 
    \label{VDM::tadpoles}
\end{align}
which allow the scalar mass matrix to be expressed as
\begin{equation}
	\mathcal{M}_{\Phi_h \Phi_S} =
	\begin{pmatrix}
		2 \lambda_H v^2 & \kappa v v_S\\
		\kappa v v_S & 2 \lambda_S v_S^2
	\end{pmatrix}
	+
	\begin{pmatrix}
		\frac{T_{\Phi_H}}{v} & 0\\
		0 & \frac{T_{\Phi_S}}{v_S}
    \end{pmatrix}\,.
    \label{VDM::massmatrix}
\end{equation}
The treatment of the tadpole contributions in the mass matrix will be
discussed in Section~\ref{sec:Renorm} while describing the
renormalisation of the tadpoles. The mass eigenstates $h_1$ and $h_2$
are obtained through the rotation with the orthogonal matrix
$R_\alpha$ as
\begin{equation}
    \begin{pmatrix}
        h_1 \\ h_2
    \end{pmatrix}
    =
    R_{\alpha} \begin{pmatrix}
        \Phi_H \\ \Phi_S
    \end{pmatrix}
    \equiv
    \begin{pmatrix}
        \cos\alpha & \sin\alpha \\
        -\sin\alpha & \cos\alpha
    \end{pmatrix}
    \begin{pmatrix}
        \Phi_H \\ \Phi_S
    \end{pmatrix} \;.
    \label{VDM::massbasis} 
\end{equation}
The diagonalisation of the mass matrix yields the mass values
$m_{h_1}$ and $m_{h_2}$ of the two scalar mass eigenstates. 
The mass of the VDM particle will be denoted as $m_{\chi}$.  
The parameters of the potential \cref{VDM::potential} can then be
expressed in terms of the \textit{physical} parameters 
\begin{equation}
    m_{h_1}\,,m_{h_2}\,,\mX\,,\alpha\,,v\,,\gX\,, T_{\Phi_H}\,, T_{\Phi_S}\,,
\end{equation}
using the relations 
\beq
\label{VDM::treelevelrelations1}
\lambda_H &=& \frac{m_{h_1}^2\cos^2\alpha+m_{h_2}^2 \sin^2\alpha}{2 v^2}\,, \\
\kappa &=& \frac{\left(m_{h_1}^2-m_{h_2}^2\right)\cos\alpha\sin\alpha}{v
  v_S}\,, \label{VDM::treelevelrelations2}\\
\lambda_S &=& \frac{m_{h_1}^2\sin^2\alpha+m_{h_2}^2 \cos^2\alpha}{2 v_S}\,,
\label{VDM::treelevelrelations3}\\
v_S &=&\frac{\mX}{\gX}  \label{VDM::treelevelrelations4} \,.
\eeq
The SM VEV $v \approx 246$~GeV is fixed by the $W$ boson mass. 
The mixing angle $\alpha$ can be chosen without loss of generality to be 
\begin{equation}
    \alpha \in \left[-\frac{\pi}{2},\frac{\pi}{2}\right)\,.
\end{equation}
The requirement of the potential to be bounded from below is
translated into the following conditions 
\beq
\lambda_H > 0, \ \ \lambda_S >0, \ \ \kappa > -2 \sqrt{\lambda_H \lambda_S}.
%\label{positivity}
\eeq
%%%%%%%%%%%%%%%%%%%%%%%%%%%%%%%%%%%%%%%%%%%%%%%%%%%%%%%
\section{Renormalisation of the VDM Model \label{sec:Renorm}}
%%%%%%%%%%%%%%%%%%%%%%%%%%%%%%%%%%%%%%%%%%%%%%%%%%%%%%%
In order to calculate the electroweak (EW) corrections to the scattering process of
the VDM particle with a nucleon we need to renormalise the VDM
model. There are four new independent parameters relative to the SM
that need to be renormalised. 
We choose them to be the non-SM-like scalar mass, $m_{h_2}$, the
rotation angle $\alpha$, the coupling  $\gX$ and the DM mass
$\mX$.\footnote{Note that in our notation $h_1$ corresponds to the
    SM-like Higgs boson, while we attribute $h_2$ to the non-SM-like scalar.}
In the following, we will present the renormalisation
of the VDM model including the gauge and Higgs sectors. \s

Having chosen the complete set of free parameters in the theory, we
start by replacing the bare parameters $p_0$ with the renormalised
ones $p$ according to  
\begin{equation}
    p_0=p + \delta p\,,
\end{equation}
where $\delta p$ is the counterterm for the parameter $p$. Denoting a
general scalar or vector field by $\Psi$, the renormalised field is
expressed in terms of the field renormalisation constant $Z_\Psi$ as 
\begin{equation}
    \Psi_0 = \sqrt{Z_{\Psi}} \Psi\,,
    \label{renorm::Zfactorgen}
\end{equation}
where $\Psi_0$ stands for the bare and $\Psi$ for the renormalised
field, respectively. In case of mixing field components, $\sqrt{Z_{\Psi}}$ is
a matrix. \s

\emph{\textit{Gauge Sector:}} Since we have an extended gauge sector
compared to the SM we will give all counterterms explicitly. Due to
the imposed $\nZ_2$ symmetry under which only the dark gauge boson
$\chi_{\mu}$ is odd, kinetic mixing between the gauge bosons of the 
$U(1)_{\X}$ and to $U(1)_Y$ is not possible. This means that there
is no interaction between the gauge sector of the SM and the new dark
gauge sector. Since this symmetry is only broken spontaneously, gauge
bosons from the two sectors will  not mix at any order of perturbation
theory and therefore the field renormalisation constants are defined
independently in each sector. We choose to renormalise the theory in
the mass basis. The replacement of the parameters in the two gauge
sectors reads  
\begin{subequations}
    \begin{align}
        &m_W^2 \rightarrow m_W^2 + \delta m_W^2\,,\\
        &m_Z^2 \rightarrow m_Z^2 + \delta m_Z^2\,,\\
        &\mX^2 \rightarrow \mX^2 + \delta \mX^2\,,\\
        &e\rightarrow  e + \delta Z_e\, e	\,,\\
        &g \rightarrow g + \delta g\,,\\
        &\gX \rightarrow \gX + \delta \gX\,,
    \end{align}
\end{subequations}
where $m_W$ and $m_Z$ are the masses of the electroweak charged
and neutral gauge bosons $W^\pm$ and $Z$, respectively, $e$ is the
electric coupling constant, and $g$ the weak $SU(2)$ coupling. 
The gauge boson fields are renormalised by their field renormalisation
constants $\delta Z$, 
\begin{subequations}
\begin{align}
     \chi &\rightarrow \cbrak{1+\frac{1}{2}\delta Z_{\chi\chi}} \chi\,,\\
     W^{\pm} &\rightarrow \cbrak{1+\frac{1}{2}\delta Z_{WW}} W^{\pm}\,,\\
    \begin{pmatrix}
        Z \\ \gamma
    \end{pmatrix} 
    &\rightarrow 
    \begin{pmatrix}
        1+\frac{1}{2} \delta Z_{ZZ} & \frac{1}{2} \delta Z_{Z\gamma}\\
        \frac{1}{2} \delta Z_{\gamma Z} & 1+\frac{1}{2} \delta Z_{\gamma\gamma}
    \end{pmatrix}
    \begin{pmatrix}
        Z \\ \gamma
    \end{pmatrix} \,.
\end{align}
\end{subequations}
The on-shell (OS) conditions yield the following expressions for the
mass counterterms of the gauge sector 
\begin{equation}
    \delta m_W^2 = \Re~\Sigma_{WW}^{T}\cbrak{m_W^2}\,,\quad  \delta
    m_Z^2 = \Re~\Sigma_{ZZ}^{T}\cbrak{m_Z^2}\,\quad\text{and}\quad
    \delta \mX^2 = \Re~\Sigma_{\chi\chi}^{T}\cbrak{\mX^2}\,, 
\end{equation}
where $T$ denotes the transverse part of the self-energies
$\Sigma_{ii}$ ($i=W,Z,\chi$). Expressing the electric charge in
terms of the Weinberg angle $\theta_W$  
\begin{equation}
    e = g \sin\theta_W, \qquad \text{with}\qquad \cos\theta_W = \frac{m_W}{m_Z},
\end{equation}
and using OS conditions for the renormalisation of the
electric charge allows for the determination of the counterterm $\delta g$
in terms of the mass counterterms $\delta m_W$, $\delta m_Z$ and
$\delta Z_e$\footnote{We use the shorthand notation
  $\sin\theta_W=s_W$ and $\cos\theta_W = c_W$.} 
\begin{align}
    \delta Z_e &=
                 \frac{1}{2}\frac{\partial\Sigma_{\gamma\gamma}^T(p^2)}{\partial
                 p^2}\bigg\vert_{p^2=0} +
                 \frac{s_W}{c_W}\frac{\Sigma_{\gamma
                 Z}^T(0)}{m_Z^2}\,,\\ 
    \frac{\delta g}{g} & = \delta Z_e +
                         \frac{1}{2}\frac{1}{m_Z^2-m_W^2}\cbrak{\delta
                         m_W^2 - c_W^2 \delta m_Z^2}\,. 
\end{align}
The wave function renormalisation constants guaranteeing the correct
OS properties are given by  
\begin{equation}
\delta Z_{\chi\chi} =
    -\Re\, \frac{\partial\Sigma_{\chi\chi}^2(p^2)}{\partial
      p^2}\bigg\vert_{p^2=m_\chi^2} \,\quad\text{,}\quad 
    \delta Z_{WW} = -\Re \,\frac{\partial\Sigma_{WW}^2(p^2)}{\partial
      p^2}\bigg\vert_{p^2=m_W^2} \,, 
\end{equation}
\begin{equation}
    \begin{pmatrix}
        \delta Z_{ZZ} & \delta Z_{Z\gamma}\\
        \delta Z_{\gamma Z} & \delta Z_{\gamma\gamma} 
    \end{pmatrix}
    = 
    \begin{pmatrix}
        -\Re\, \frac{\partial \Sigma_{ZZ}^T(p^2)}{\partial
          p^2}\bigg\vert_{p^2=m_Z^2}  & 2
        \frac{\Sigma_{Z\gamma}^T(0)}{m_Z^2}\\ 
        -2 \frac{\Sigma_{Z\gamma}^T(0)}{m_Z^2} & -\Re \,\frac{\partial
          \Sigma_{\gamma\gamma}^T(p^2)}{\partial p^2}\bigg\vert_{p^2=0} 
    \end{pmatrix}\,.
\end{equation}
As for the gauge coupling from the dark sector, $\gX$, since there is
no obvious physical quantity to fix the renormalisation constant, we
will renormalise it using the $\overline{\mbox{MS}}$ scheme, which will be
described in detail in Section~\ref{sec::gX}. \s

\emph{\textit{Higgs Sector:}}
In the VDM model we have two scalar fields which mix, namely the real component 
$\Phi_H$ of the Higgs doublet and the real component $\Phi_S$ of the
singlet, yielding the mass eigenstates $h_1$ and $h_2$. This mixing
has to be accounted for in the field renormalisation constants (see
\cref{renorm::Zfactorgen}) so that the corresponding
matrix reads  
\begin{equation}
    \begin{pmatrix}
        h_1 \\ h_2
    \end{pmatrix}
    \rightarrow
    \begin{pmatrix}
        1+\frac{1}{2}\delta Z_{h_1 h_1} & \frac{1}{2}\delta Z_{h_1 h_2}\\
        \frac{1}{2} \delta Z_{h_2 h_1} & 1+\frac{1}{2} \delta Z_{h_2 h_2}
    \end{pmatrix}
    \begin{pmatrix}
        h_1 \\ h_2
    \end{pmatrix}\,.
    \label{renorm::ZfactorHiggs}
\end{equation}
In the mass eigenbasis, the mass matrix in \cref{VDM::massmatrix} yields
\begin{equation}
    \mathcal{M}_{h_1 h_2} = 
    \underbrace{
    \begin{pmatrix}
        m_{h_1}^2 & 0\\
        0 & m_{h_2}^2
    \end{pmatrix}
    }_{\equiv \mathcal{M}^2}
    +
    \underbrace{
    R_{\alpha} \begin{pmatrix}
        T_{\Phi_H}/v & 0 \\
        0& T_{\Phi_S}/v_S
    \end{pmatrix}
    R_{\alpha}^T
    }_{\equiv\delta T}\,.
    % \equiv 
    % \begin{pmatrix}
    %     \mh^2 & 0\\
    %     0 & \mH^2
    % \end{pmatrix}
    % + \
    \label{RENORM::MASSMATRIX}
\end{equation}
The tadpole terms in the tree-level mass matrix are \textit{bare
  parameters}. At next-to-leading order (NLO) they obtain a shift
that corresponds to the 
change of the vacuum state of the potential through electroweak
corrections. To avoid such vacuum shifts at NLO, we renormalise the
tadpoles such that the VEV remains at its tree-level value also at
NLO. This requires the introduction of tadpole counterterms $\delta
T_i$ such that the one-loop renormalised one-point $\hat T_i$ function
vanishes 
\begin{equation}
    \hat T_i =T_i - \delta T_i \overset{!}{=}0\,,\quad
    i =  \Phi_H , \Phi_S \,. 
\end{equation}
Since we formulate all counterterms in the mass basis it is convenient
to rotate the tadpole parameters in their corresponding mass
basis as well, using the same rotation matrix $R_\alpha$, 
\begin{equation}
    \begin{pmatrix}
        T_{h_1}\\T_{h_2}
    \end{pmatrix}
    =
    R_{\alpha}\cdot 
    \begin{pmatrix}
    T_{\Phi_h}\\T_{\Phi_S}
    \end{pmatrix}\,.
    \label{RENORM::TadpolCondi}
\end{equation}
For the mass counterterms of the Higgs sector we replace the mass matrix as
\begin{equation}
    \mathcal{M}_{h_1 h_2} \rightarrow \mathcal{M}_{h_1 h_2}+\delta
    \mathcal{M}_{h_1 h_2} \;,
\end{equation}
with the one-loop counterterm 
\begin{equation}
    \delta \mathcal{M}_{h_1 h_2} =
    \begin{pmatrix}
    \delta m_{h_1}^2 & 0 \\
    0&\delta m_{h_2}^2    
    \end{pmatrix}
    + R_{\alpha} 
    \begin{pmatrix}
        \frac{\delta T_{\Phi_H}}{v} & 0 \\
        0 & \frac{\delta T_{\Phi_S}}{v_S}
    \end{pmatrix}
    R_{\alpha}^T
    \equiv 
    \begin{pmatrix}
        \delta m_{h_1}^2 & 0 \\
        0&\delta m_{h_2}^2 
        \end{pmatrix}
    +  \begin{pmatrix}
        \delta T_{h_1 h_1} & \delta T_{h_1 h_2} \\
        \delta T_{h_2 h_1}&\delta T_{h_2 h_2}    
        \end{pmatrix}\,.
        \label{RENORM::MASSCOUNTERTERM}
 \end{equation}
In \cref{RENORM::MASSCOUNTERTERM} we neglect all terms of order
$\mathcal O \cbrak{\delta\alpha\delta T_i}$ since they
are formally of two-loop order. Using OS conditions and
\cref{RENORM::MASSCOUNTERTERM} finally yields the
following relations for the counterterms ($i=1,2$) 
\begin{align}
    &\delta m^2_{h_i} = \Re\sbrak{\Sigma_{h_ih_i}(m_{h_i}^2) - \delta T_{h_ih_i}}\,,\\
    &\delta Z_{h_ih_i}  = -
      \Re\sbrak{\frac{\partial\Sigma_{h_ih_i}(p^2)}{\partial
      p^2}}_{p^2=m_{h_i}^2}\,,\\ 
    &\delta Z_{h_ih_j} = \frac{2}{m_{h_i}^2-m_{h_j}^2}
      \Re\sbrak{\Sigma_{h_ih_j}(m_{h_j}^2)-\delta T_{h_ih_j}}\,,\quad
      i\neq j\,.\label{RENORM::ZFACTOR} 
\end{align}

%%%%%%%%%%%%%%%%%%%%%%%%%%%%%%%%%%%%%%%%%%%%%%%%%%%%%%%%%%%%%
\subsection{Renormalisation of the Dark Gauge Coupling
  $\gX$}\label{sec::gX}
%%%%%%%%%%%%%%%%%%%%%%%%%%%%%%%%%%%%%%%%%%%%%%%%%%%%%%%%%%%%%
As previously mentioned, the dark gauge coupling $\gX$ cannot be linked to
a physical observable, which prevents the usage of OS conditions for
its renormalisation. Therefore the coupling will be renormalised using the
$\overline{\mbox{MS}}$ scheme. As 
the UV divergence is universal, we just need a vertex involving $\gX$.
We choose the triple $h_1h_1h_1$ vertex. First we write 
\begin{equation}
    \mathcal A^{\text{NLO}}_{h_1h_1h_1} = \mathcal A^{\text{LO}}_{h_1h_1h_1} + \mathcal A^{\text{VC}}_{h_1h_1h_1} +\mathcal A^{\text{CT}}_{h_1h_1h_1}\,,
\end{equation}
where $\mathcal{A}^{\text{VC}}$ stands for the amplitude for the
virtual corrections to the vertex and $\mathcal A^{\text{CT}}$ is the
amplitude for the vertex counterterm. We will henceforth drop the
index  $h_1h_1h_1$ for better readability. 
 The counterterm amplitude $\mathcal A^{\text{CT}}$ consists of two contributions,
\begin{equation}
    \mathcal{A}^{\text{CT}} = \delta^{\text{mix}} + \delta g^{\text{CT}}
\end{equation}
with 
\begin{equation}
    \delta^{\text{mix}} = \frac{3}{2} g_{h_1h_1h_1} \delta Z_{h_1h_1} + \frac{3}{2} g_{h_1h_1h_2} \delta Z_{h_2h_1}    
\end{equation}
and 
\begin{equation}
    \delta g^{\text{CT}} = \sum_p\frac{\partial
      g_{h_1h_1h_1}}{\partial p} \delta p\,,\quad p\in
\lbrace \mh,\mH,m_\chi,v,\alpha,\gX\rbrace\,. 
\end{equation}
The trilinear Higgs self-coupling reads (expressing $v$ through $2m_W/g$)
\beq
g_{h_1 h_1 h_1 } = - \frac{3 g m_{h_1}^2}{2 m_W} \cos^3\alpha - \frac{3 
\gX m_{h_1}^2}{\mX} \sin^3\alpha \;.
\eeq
The divergent part of $ \delta \gX$ is then given by 
\begin{equation}
    \delta \gX\big\vert_{\text{div}} = \cbrak{\frac{\mX}{3 \mh^2\sin^3\alpha}} \cbrak{\mathcal A^{\text{VC}}+\mathcal{A}^{\text{CT}}\big\vert_{\delta \gX =0}}\big\vert_{\text{div}} \,.
    \label{renorm::CTgX}
\end{equation}

\begin{figure}[ht!]
    \centering
    \footnotesize
    \stackunder[5pt]{\includegraphics[width=0.15\textwidth]{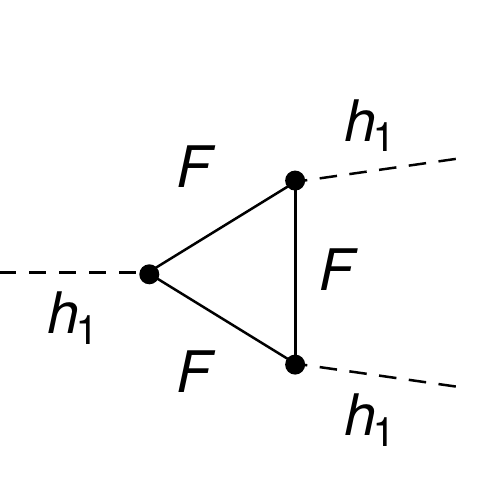}}{\tiny $F=\lbrace l ,q\rbrace$}%
    \hspace{0.25cm}%
    \stackunder[5pt]{\includegraphics[width=0.15\textwidth]{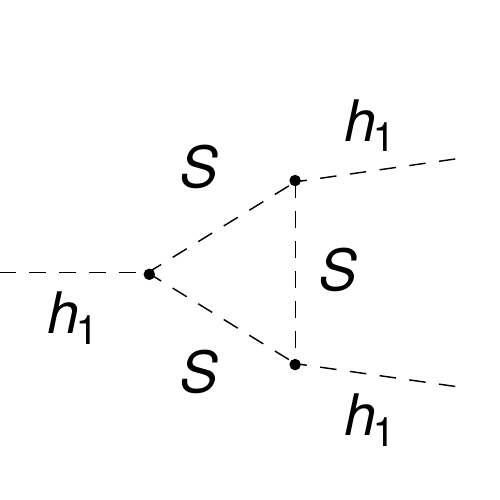}}{\tiny$S=\lbrace h_i,G^0,G^{\pm},G_{\chi} \rbrace$}
    \hspace{0.25cm}%
    \stackunder[5pt]{\includegraphics[width=0.15\textwidth]{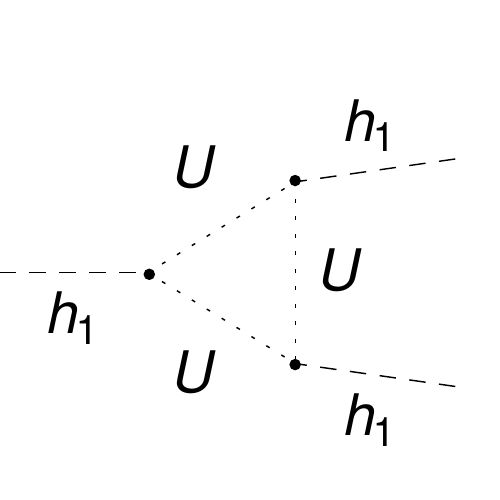}}{\tiny$U\in\lbrace \eta_Z,\eta_{\pm},\eta_{\chi}\rbrace$}
    \hspace{0.25cm}%
    \stackunder[5pt]{\includegraphics[width=0.15\textwidth]{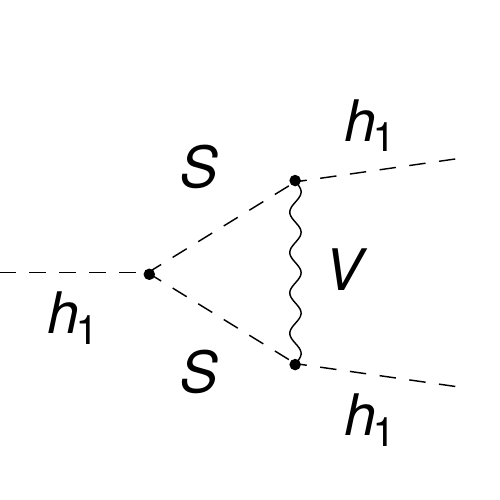}}{\tiny$S,V=\lbrace G^0,G^{\pm},G_{\chi}\rbrace,\lbrace Z,W^\pm,X\rbrace$}
    \hspace{0.25cm}%
    \stackunder[5pt]{\includegraphics[width=0.15\textwidth]{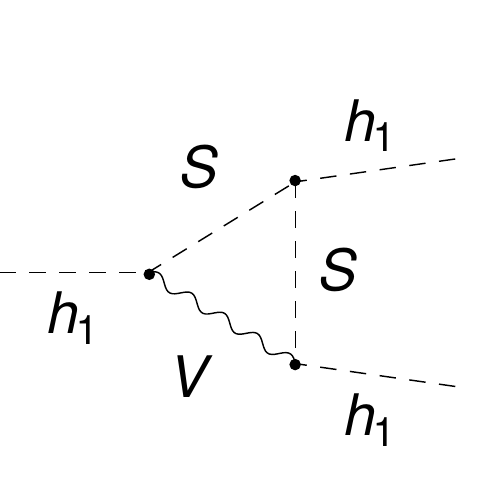}}{\tiny$S,V=\lbrace G^0,G^{\pm},G_{\chi}\rbrace,\lbrace Z,W^\pm,X\rbrace$}
    \hspace{0.25cm}%
    \stackunder[5pt]{\includegraphics[width=0.15\textwidth]{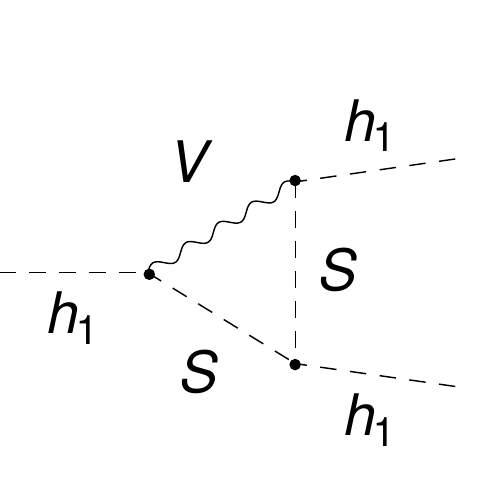}}{\tiny$S,V=\lbrace G^0,G^{\pm},G_{\chi}\rbrace,\lbrace Z,W^\pm,X\rbrace$}
    \hspace{0.25cm}%
    \stackunder[5pt]{\includegraphics[width=0.15\textwidth]{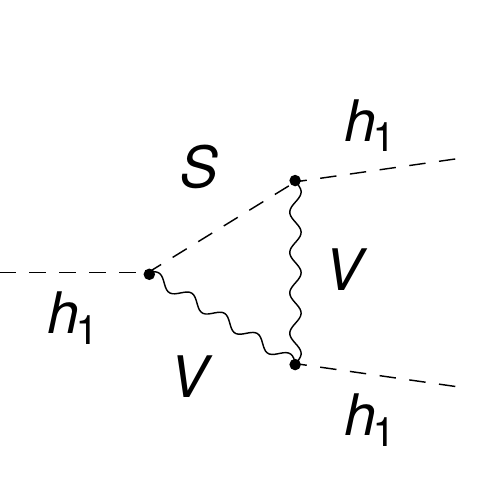}}{\tiny$S,V=\lbrace G^0,G^{\pm},G_{\chi}\rbrace,\lbrace Z,W^\pm,X\rbrace$}
    \hspace{0.25cm}%
    \stackunder[5pt]{\includegraphics[width=0.15\textwidth]{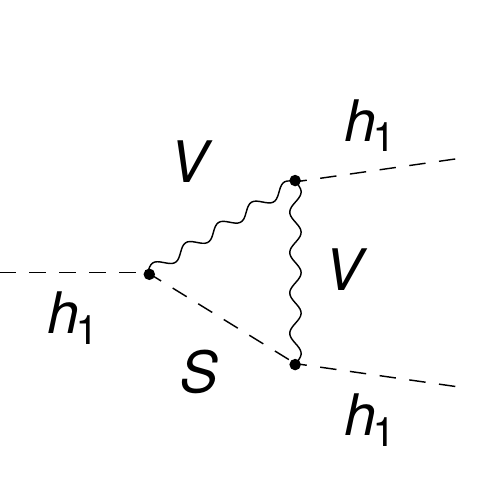}}{\tiny$S,V=\lbrace G^0,G^{\pm},G_{\chi}\rbrace,\lbrace Z,W^\pm,X\rbrace$}
    \hspace{0.25cm}%
    \stackunder[5pt]{\includegraphics[width=0.15\textwidth]{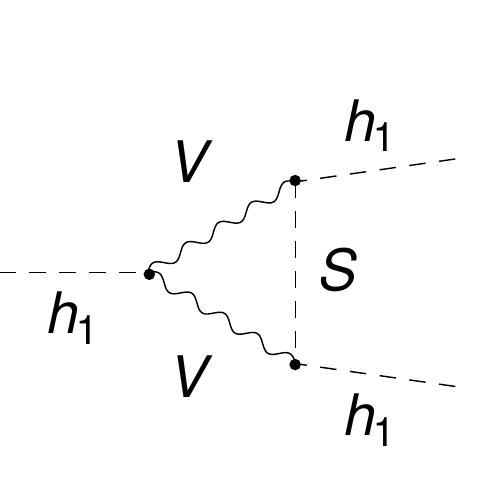}}{\tiny$S,V=\lbrace G^0,G^{\pm},G_{\chi}\rbrace,\lbrace Z,W^\pm,X\rbrace$}
    \hspace{0.25cm}%
    \stackunder[5pt]{\includegraphics[width=0.15\textwidth]{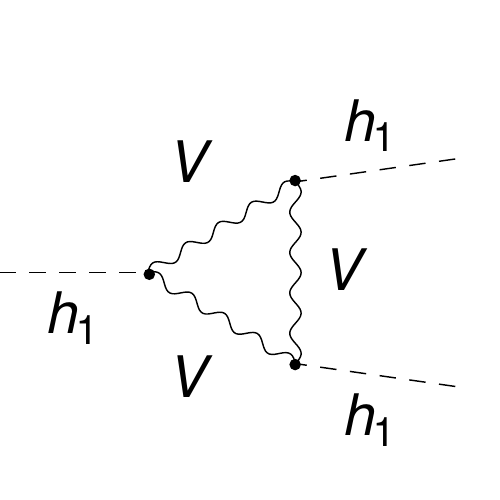}}{\tiny$V=\lbrace Z,W^\pm,X\rbrace$}\\
    \hspace{0.25cm}%
    \stackunder[5pt]{\includegraphics[width=0.15\textwidth]{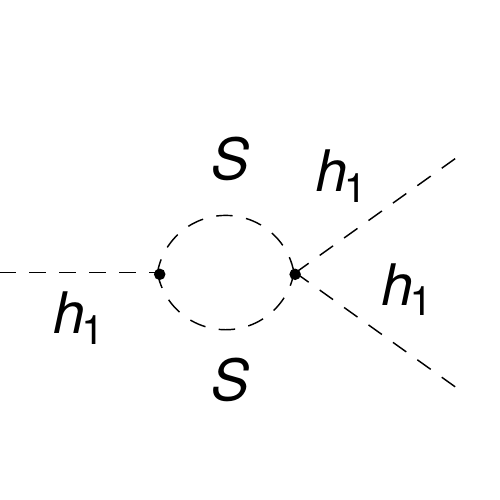}}{\tiny$S=\lbrace G^0,G^{\pm},G_{\chi},h_i\rbrace$}
    \hspace{0.25cm}%
    \stackunder[5pt]{\includegraphics[width=0.15\textwidth]{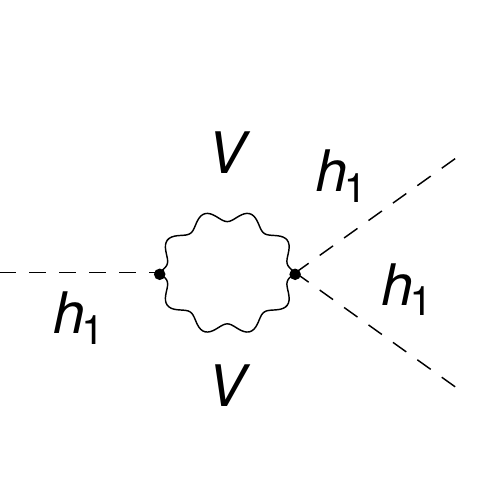}}{\tiny$V=\lbrace Z,W^\pm,X\rbrace$}
    \hspace{0.25cm}%
    \stackunder[5pt]{\includegraphics[width=0.15\textwidth]{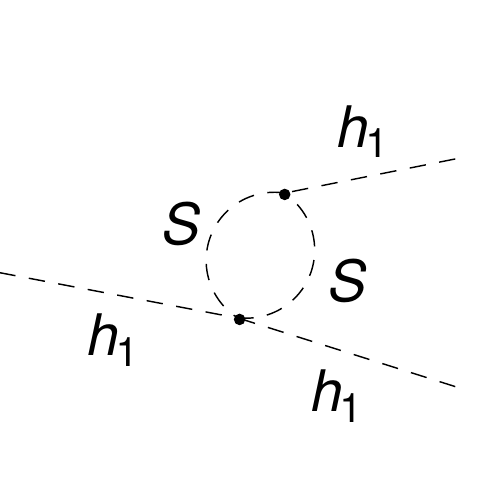}}{\tiny$S=\lbrace G^0,G^{\pm},G_{\chi},h_i\rbrace$}\\
    \hspace{0.25cm}%
    \stackunder[5pt]{\includegraphics[width=0.15\textwidth]{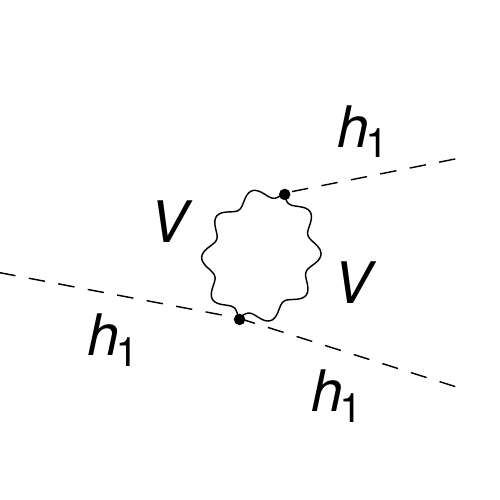}}{\tiny$V=\lbrace Z,W^\pm,X\rbrace$}
    \hspace{0.25cm}%
    \stackunder[5pt]{\includegraphics[width=0.15\textwidth]{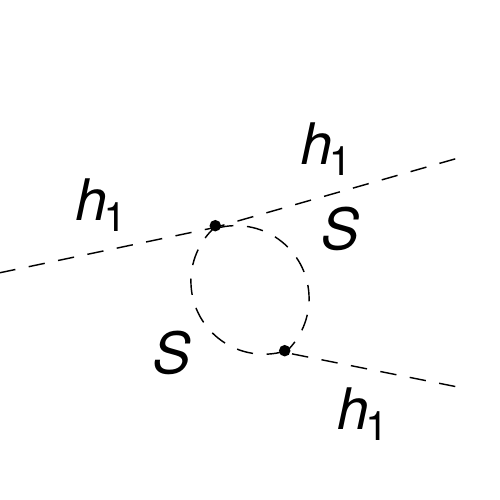}}{\tiny$S=\lbrace G^0,G^{\pm},G_{\chi},h_i\rbrace$}
    \hspace{0.25cm}%
    \stackunder[5pt]{\includegraphics[width=0.15\textwidth]{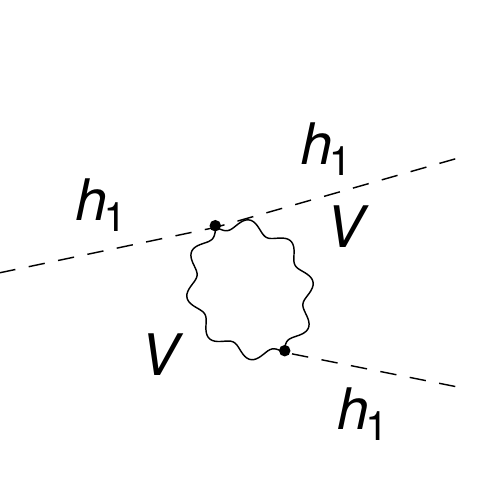}}{\tiny$V=\lbrace Z,W^\pm,X\rbrace$}

    \caption{Generic diagrams contributing to $\mathcal
    A^{\text{VC}}_{h_1h_1h_1}$. Here $F$ denotes fermions, $S$ scalars, 
    $V$ gauge bosons, and $U$ ghost fields.}
    \label{renorm::VC}
\end{figure}

In \cref{renorm::VC} we present the set of diagrams used to calculate
$\mathcal A^{\text{VC}}$. The one-loop diagrams were generated with {\tt
  FeynArts}\cite{Hahn:2000kx} for which the model file was obtained with {\tt SARAH} 
\cite{Staub:2013tta,Staub:2012pb,Staub:2010jh,Staub:2009bi} and the
program package {\tt FeynCalc}
\cite{Shtabovenko:2016sxi,Mertig:1990an} was used to reduce the
amplitudes to Passarino-Veltmann integrals
\cite{Passarino:1978jh}. The numerical evaluation of the integrals was
done by {\tt Collier}
\cite{Denner:2016kdg,Denner:2002ii,Denner:2005nn,Denner:2010tr}. 
The counterterm $\gX$ in the $\overline{\mbox{MS}}$ scheme is then obtained as
\begin{equation}
    \delta \gX\big\vert_{\varepsilon} = \frac{\gX^3}{96\pi^2}\Delta_{\varepsilon}\,,
\end{equation}
with 
\beq
\Delta_{\varepsilon} = \frac{1}{\varepsilon} - \gamma_E +\ln
4\pi \;,
\eeq
where $\gamma_E$ denotes the Euler-Mascheroni constant. 

%%%%%%%%%%%%%%%%%%%%%%%%%%%%%%%%%%%%%%%%%%%%%%%%%%%%%%%%%%%%%%
\subsection{Renormalisation of the Scalar Mixing Angle
  $\alpha$ \label{sec:alpharenorm}} 
%%%%%%%%%%%%%%%%%%%%%%%%%%%%%%%%%%%%%%%%%%%%%%%%%%%%%%%%%%%%%%
The final parameter that needs to be renormalised is the mixing angle
$\alpha$. Again, this is a quantity that cannot be related directly to
an observable, except if we would use a process-dependent
renormalisation scheme which is known to lead to unphysically large
counterterms~~\cite{Krause:2016oke}. The renormalisation 
of the mixing angles in SM extensions was thoroughly discussed
in~\cite{Bojarski:2015kra,Krause:2016oke,Denner:2016etu,Krause:2016xku,Krause:2017mal,Altenkamp:2017ldc,Altenkamp:2017kxk,Fox:2017hbw,
  Denner2018, Grimus:2018rte,Krause:2018wmo,Krause:2019oar}. In this
work we will use the  
KOSY scheme, proposed in~\cite{Pilaftsis:1997dr, Kanemura:2004mg}, which
connects for the derivation of the angle counterterm the usual OS conditions of the 
scalar field with the relations between the gauge basis and the mass
basis. The bare parameter expressed through the renormalised one and
the counterterm reads
\begin{equation}
    \alpha_0 = \alpha + \delta \alpha\,.
\end{equation}
Considering the field strength renormalisation before the  rotation, 
\begin{equation}
    \begin{pmatrix}
        h_1\\h_2
    \end{pmatrix}
    = R\cbrak{\alpha+\delta\alpha} \sqrt{Z_{\Phi}}\begin{pmatrix}
        \Phi_H\\\Phi_S
    \end{pmatrix}\,,
\end{equation}
and expanding it to strict one-loop order, 
\begin{equation}
    R\cbrak{\alpha+\delta\alpha} \sqrt{Z_{\Phi}}
    \begin{pmatrix}
        \Phi_H\\\Phi_S
    \end{pmatrix}
    =\underbrace{R(\delta\alpha) R(\alpha)
      \sqrt{Z_{\Phi}}R(\alpha)^T}_{\overset{!}{=}\sqrt{Z_H}} R(\alpha)  
    \begin{pmatrix}
        \Phi_H\\\Phi_S
    \end{pmatrix}
    +\mathcal{O} (\delta\alpha^2)
    = \sqrt{Z_H}   
    \begin{pmatrix}
        h_1\\h_2
    \end{pmatrix}\,,
\end{equation}
yields the field strength renormalisation matrix $\sqrt{Z_H} $
connecting the bare and renormalised fields in the mass basis. Using
the rotation matrix expanded at one-loop order results in 
\begin{equation}
    \sqrt{Z_{H}} = R(\delta\alpha)
    \begin{pmatrix}
        1+\frac{\delta Z_{h_1h_1}}{2} & \delta C_{h}\\
        \delta C_h & 1 + \frac{\delta Z_{h_2h_2}}{2}    
    \end{pmatrix}
    \approx 
    \begin{pmatrix}
    1+\frac{\delta Z_{h_1h_1}}{2} & \delta C_h +\delta \alpha\\
    \delta C_h - \delta \alpha & 1+\frac{\delta Z_{h_2h_2}}{2}  
    \end{pmatrix}\,.
\end{equation}
Demanding that the field mixing vanishes on the mass shell is
equivalent to identifying the off-diagonal elements of
$\sqrt{Z_H}$ with those in \cref{renorm::ZfactorHiggs}, 
\begin{equation}
    \frac{\delta Z_{h_1h_2}}{2} \overset{!}{=} \delta C_h +\delta
    \alpha \qquad\text{and}\qquad \frac{\delta Z_{h_2h_1}}{2}
    \overset{!}{=} \delta C_h -\delta \alpha\,. 
\end{equation}
With \cref{RENORM::ZFACTOR} the mixing angle counterterm reads 
\begin{eqnarray}
    \delta \alpha & = & \frac{1}{4}\cbrak{\delta Z_{h_1h_2}-\delta Z_{h_2h_1}} \\
    &&=
       \frac{1}{2(m_{h_1}^2-m_{h_2}^2)}\Re\cbrak{\Sigma_{h_1h_2}(m_{h_1}^2)+
     \Sigma_{h_1h_2}(m_{h_2}^2)- 2 \delta T_{h_1h_2}}\,. 
\end{eqnarray}

In our numerical analysis we will use two more renormalisation
schemes for $\delta\alpha$: the $\overline{\mbox{MS}}$ scheme and a
process-dependent scheme. In the $\overline{\mbox{MS}}$ scheme we only
take the counterterm $\delta \alpha$ into account in the divergent
parts in $D=4$ dimensions. Applying dimensional regularisation
\cite{tHooft:1972tcz,tHooft:1973mfk}, these are the terms proportional
to $1/\epsilon$, where $D= 4 - 2\epsilon$. Both the KOSY scheme and
the $\overline{\mbox{MS}}$ scheme lead to a gauge-parameter dependent
definition of $\delta \alpha$ 
This is avoided if  $\delta \alpha$ is defined through a physical process. \s 

In our process-dependent
renormalisation scheme for $\alpha$, discussed in the numerical results, we define
the counterterm $\delta \alpha$ through the process $h \to \tau
\tau$, where $h$ denotes the SM-like scalar of the two $h_i$ ($i=1,2$). 
The counterterm is defined by requiring the NLO decay
width to be equal to the LO one.
The NLO corrections involve infrared (IR) divergences 
stemming from the QED corrections. Since they form a UV-finite subset, 
this allows us to apply the renormalisation condition solely on the
weak sector thus avoiding the IR divergences, {\it i.e.}
we require for the NLO and LO amplitudes of the decay process
\begin{equation}
\mathcal A_{h \to \tau\tau}^{\text{NLO,weak}} \overset{!}{=} \mathcal
A_{h \to \tau\tau}^{\text{LO}}\,, \label{eq:procdep}
\end{equation}
where 'weak' refers to the weak part of the NLO amplitude. 
The $h$ coupling to $\tau \bar{\tau}$ depends on the
mixing angle $\alpha$ as 
\beq
g_{h\tau\tau} = \frac{g m_\tau \cos\alpha}{2 m_W} \;,
\eeq
and the LO amplitude reads
\begin{equation}
\mathcal A_{h \to \tau\tau}^{\text{LO}} = g_{h\tau\tau} \bar
u(p_{\tau})u(p_{\tau}) = \frac{g m_{\tau}\cos\alpha}{2 m_W} \bar
u(p_{\tau})u(p_{\tau}) \,,
\end{equation}
with $u(p_\tau)$ ($\bar{u}(p_\tau)$) denoting the spinor (anti-spinor)
of the $\tau$ with four-momentum $p_\tau$. Dividing the weak NLO amplitude
into the LO amplitude, the weak virtual corrections to the amplitude,
and the corresponding counterterm part, 
\beq
{\cal A}^{\text{NLO,weak}}_{h\to \tau\tau} = {\cal A}^{\text{LO}} + {\cal
  A}^{\text{virt,weak}} + {\cal A}^{\text{ct}} \;,
\eeq
the condition Eq.~(\ref{eq:procdep}) translates into 
\beq
{\cal A}^{\text{virt,weak}} + {\cal A}^{\text{ct}} = 0 \;,
\eeq
and we get the mixing angle counterterm in the process-dependent
scheme as
\beq
\delta \alpha = \left( \frac{2m_W}{gm_\tau \cos\alpha} \right) \left[
  {\cal A}^{\text{virt,weak}} + \left. {\cal A}^{\text{ct}}\right|_{\delta
    \alpha=0} \right] \;.
\eeq 
Here $\left. {\cal A}^{\text{ct}}\right|_{\delta \alpha=0}$ denotes the
complete counterterm amplitude but without the contribution from $\delta
\alpha$.

%%%%%%%%%%%%%%%%%%%%%%%%%%%%%%%%%%%%%%%%%%%%%%%%%%%%%%%
\section{Dark Matter Direct Detection at Tree Level\label{sec:ddtree}}
%%%%%%%%%%%%%%%%%%%%%%%%%%%%%%%%%%%%%%%%%%%%%%%%%%%%%%%
In the following we want to set our notation and conventions used in
the calculation of the spin-independent (SI) cross section of
DM-nucleon scattering. The interaction between the DM and the nucleon 
is described in terms of effective coupling constants. The major
contribution to the cross section comes from the light quarks
$q=u,d,s$ and gluons. For the VDM model 
the effective operator basis contributing to the SI cross section is
given by \cite{Hisano:2010yh} 
\begin{equation}
    \mathcal L ^{\text{eff}} = \sum_{q=u,d,s}\mathcal L^{\text{eff}}_q + \mathcal
    L^{\text{eff}}_G \;,
    \label{HISANO::LEFF}
\end{equation}
with 
\begin{subequations}    
    \begin{align}
        &\mathcal L ^{\text{eff}}_q = f_q \chi_{\mu}\chi^{\mu} m_q
          \bar q q  +
          \frac{g_q}{\mX^2}\chi^{\rho}\ii \partial^{\mu}\ii\partial^{\nu}\chi_{\rho}
         \mathcal{O}^q_{\mu\nu}\,,\\  
        &\mathcal L^{\text{eff}}_G = f_G \chi_{\rho}\chi^{\rho}
          G^{a}_{\mu\nu}G^{a \,\mu\nu}\,, \label{eq:ggxx}
    \end{align}
    \label{tree::effoperators}
\end{subequations}
where $G_{\mu\nu}^a$ ($a=1,...,8$) denotes the gluon
field strength tensor and $\mathcal{O}_{\mu\nu}^q$ the quark twist-2
operator corresponding to the traceless part of the energy-momentum
tensor of the nucleon \cite{Hisano:2010ct,Hisano:2015bma}, 
\begin{equation}
    \mathcal{O}_{\mu\nu}^q = \frac{1}{2}\bar q \ii
    \cbrak{\partial_{\mu}\gamma_{\nu}+\partial_{\nu}\gamma_{\mu}-
    \frac{1}{2}\slashed{\partial}}q\,.  
\end{equation}
Operators suppressed by the DM velocities and the momentum transfer of
the DM particle to the nucleon are neglected in the analysis. Furthermore, we neglect
contributions introduced by the gluon twist-2 operator $\mathcal
O_{\mu\nu}^g$, since these contributions are one order higher in the
strong coupling constant $\alpha_s$ \cite{Hisano:2010yh}. 
\s

For vanishing momentum transfer and on-shell nucleon states, the
nucleon matrix elements are given by 
\begin{subequations}    
\beq
        \bra{N}m_q\bar q q \ket{N} &=& m_N f^N_{T_q} \\
       - \frac{9\alpha_S}{8\pi} \bra{N}G_{\mu\nu}^a
        G^{a,\mu\nu}\ket{N} &=& \left(1 - \sum_{q=u,d,s}
          f^N_{T_q}\right) m_N = m_N f^N_{T_G} \\
        \bra{N(p)}\mathcal O_{\mu\nu}^q\ket{N(p)} &=&
        \frac{1}{m_N}\cbrak{p_{\mu}p_{\nu}-\frac{1}{4}m_N^2
          g_{\mu\nu}}\cbrak{q^N(2)+\bar q^N(2)} \;,
\eeq
    \label{HISANO::MATRIXELEMENT}
\end{subequations}
where $N$ denotes a nucleon, $N=p,n$, and $m_N$ is the 
nucleon mass. Furthermore, $q^N(2), \bar q^N(2)$ are the second
moments of the parton distribution functions of the quark $q(x)$ and
the antiquark $\bar q(x)$, respectively. The four-momentum
  of the nucleon is denoted by $p$. The numerical values for the
matrix elements $f^N_{T_q}$, $f^N_{T_G}$ and the second moments
$q^N(2)$ and $\bar{q}^N(2)$ are given in App.~\ref{APP::NUCLEAR}. The SI effective
coupling of the VDM particle with the nucleons is obtained from the
nucleon expectation value of the effective Lagrangian,
Eq.~(\ref{HISANO::LEFF}), by applying
Eqs.~(\ref{HISANO::MATRIXELEMENT}), which yields
\begin{equation}
    f_N/m_N = \sum_{q=u,d,s} f_q f^N_{T_q} + \sum_{q=u,d,s,c,b}
    \frac{3}{4}\cbrak{q^N(2)+\bar q^N(2)} g_q -\frac{8\pi}{9\alpha_S}
    f^N_{T_G}f_G\,.
\label{eq:fnovermn}
\end{equation}
In the contribution from the quark twist-2 operator all quarks
below the energy scale $\sim 1$~GeV have to be included,
{\it i.e.}~all quarks but the top quark.
The SI scattering cross section between the VDM particle and a
nucleon, proton or neutron ($N=p,n$), is given by  
\begin{equation}
    \sigma_N = \frac{1}{\pi}\cbrak{\frac{m_N}{\mX+m_N}}^2\big\vert f_N\big\vert^2\,.
    \label{TREE::CX}
\end{equation}
\begin{figure}
    \centering
    \includegraphics[width=0.2\textwidth]{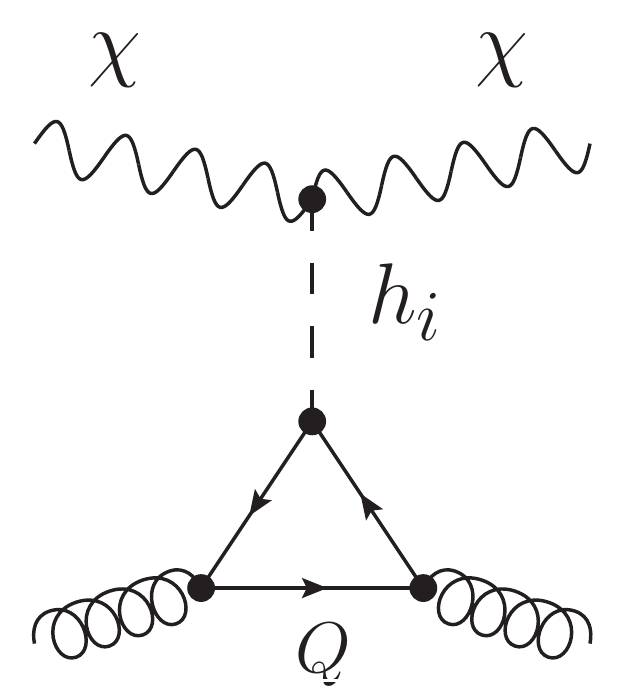}
\vspace*{0.2cm}
    \caption{Higgs bosons $h_i$ mediating the coupling of two
      gluons to two VDM particles through a heavy
      quark loop. \label{fig:gluontriangle}} 
\end{figure} 
Note that the sum in the first term of Eq.~(\ref{eq:fnovermn}) only
extends over the light quarks. The leading-order gluon interaction
with two VDM particles is mediated by one of the two Higgs bosons
which couple to the external gluons through a heavy quark triangle diagram, {\it
  cf.}~Fig.~\ref{fig:gluontriangle}. The charm, bottom and top quark
masses are larger than the energy scale relevant for DM direct
detection and should therefore be integrated out for the description
of the interaction at the level of the nucleon.  By calculating the
heavy quark triangle diagrams and then integrating out the heavy
quarks we obtain the related operator in the heavy quark limit. This
is equivalent to calculating the diagram in Fig.~\ref{fig:trelevdiag}
with heavy quarks $Q = c,b,t$, and replacing the resulting tensor
structure $m_Q  \bar{Q}Q$ with the effective gluon operator as follows
\cite{Shifman:1978zn,Ertas:2019dew,Abe:2018emu} 
\begin{equation}
    m_Q \bar Q Q \rightarrow - \frac{\alpha_S}{12\pi} G_{\mu\nu}^aG^{a\mu\nu}\,,
    \label{tree::gluonmap}
\end{equation}
corresponding to the effective leading-order VDM-gluon interaction in
\cref{tree::effoperators}. \s

For the tree-level contribution to the SI cross section the
$t$-channel diagrams depicted in Fig.~\ref{fig:trelevdiag} have to be
calculated for vanishing momentum 
transfer. The respective Wilson coefficient for the effective operator
in \cref{HISANO::LEFF} is extracted by projecting 
onto the corresponding tensor structure, $m_q q \bar{q}$. Accounting
for the additional symmetry factor 
of the amplitude, this yields finally the following $f_q$
factor for the quarks,
\begin{equation}
    f_q = \frac{1}{2} \frac{g
      \gX}{m_W}\frac{\sin(2\alpha)}{2}\frac{\mh^2-\mH^2}{\mh^2\mH^2}
    \mX\,,\quad q=u,d,s,c,b,t\,.
\label{eq:fqlo}
\end{equation}
As explained above, the heavy quarks $Q=b,c,t$ contribute to the effective gluon interaction. By using
Eq.~(\ref{tree::gluonmap}), the Wilson coefficient for the gluon
interaction, $f_G$, can be expressed in 
terms of $f_q$ for $q=c,b,t$,
\beq
f_G = \sum_{q=c,b,t} - \frac{\alpha_S}{12\pi} f_q \;,
\eeq
resulting in the SI LO cross section
\begin{equation}
    \sigma^{\text{LO}} = \frac{\sin^22\alpha}{4\pi}\cbrak{\frac{\mX
        m_N}{\mX+m_N}}^2\frac{\cbrak{\mh^2-\mH^2}^2}{\mh^4\mH^4}\,\,\frac{\mX^2
      m_N^2}{v^2 v_S^2}\,\left| \sum_{q=u,d,s}
    f^N_{T_q}+3 \cdot \frac{2}{27}f^N_{T_G}\right|^2\,. \label{eq:siddlo}
\end{equation}
The twist-2 operator does not contribute at LO. The obtained result is in agreement with
Ref.~\cite{Azevedo:2018oxv}\footnote{The authors of Ref.~\cite{Azevedo:2018oxv} 
  introduced an effective coupling $f_N\approx 0.3$ between the
  nucleon and the DM particle, which corresponds to $\vert \sum_{q=u,d,s}
  f_{T^N_q}+\frac{2}{9}f_{T^N_G}\big\vert$.}.
\begin{figure}
    \centering
    \includegraphics[width=0.2\textwidth]{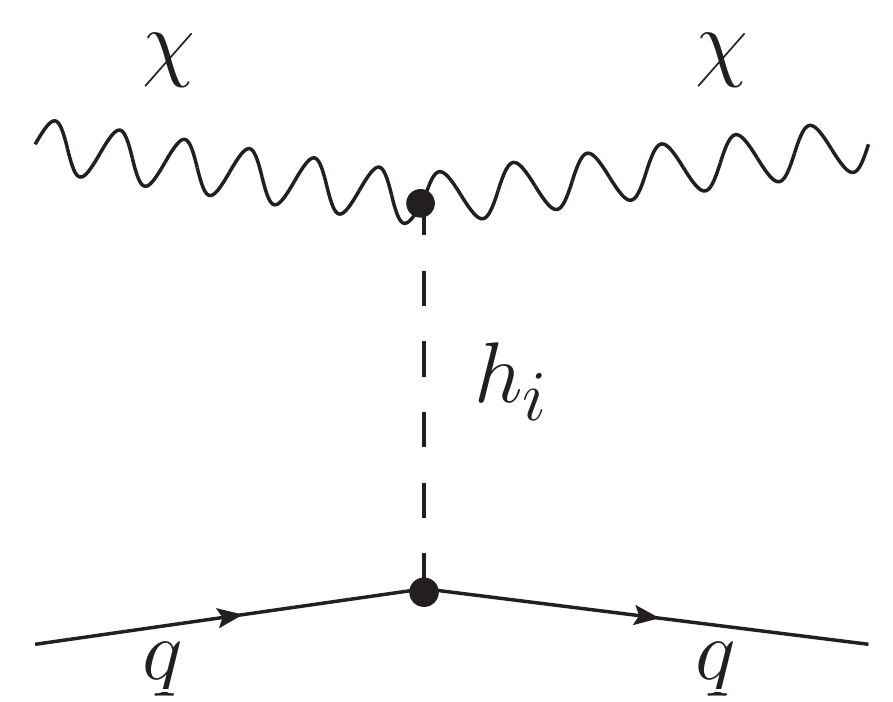}
\vspace*{0.2cm}
    \caption{Generic tree-level diagram contribution to the SI cross
      section. The mediator $S$ corresponds to the two Higgs bosons
      $h_1$ and $h_2$. The quark line $q$ corresponds to all quarks
      $q=u,d,s,c,b,t$. \label{fig:trelevdiag}} 
\end{figure}   

%%%%%%%%%%%%%%%%%%%%%%%%%%%%%%%%%%%%%%%%%%%%%%%%%%%%%%%%%%%%%
\section{Dark Matter Direct Detection at One-Loop
  Order\label{sec:dd1loop}} 
\begin{figure}
    \centering
    \subfigure[Vertex
    Corrections\label{VERTEXCORRECTIONS}]{\includegraphics[width=0.25\textwidth]{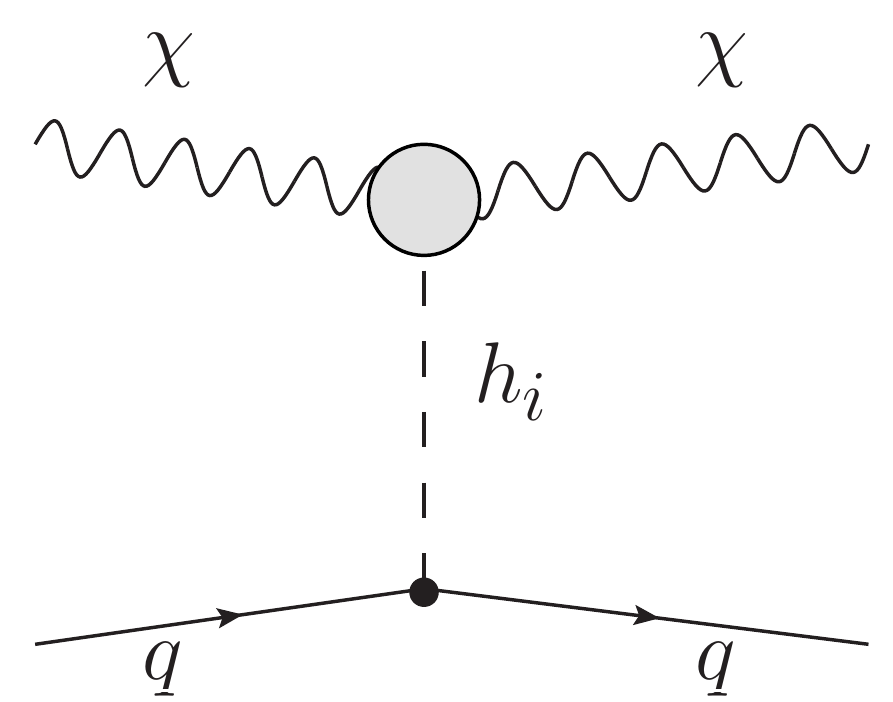}} \hspace*{0.4cm}
    \subfigure[Mediator
    Corrections\label{MEDIATORCORRECTIONS}]{\includegraphics[width=0.25\textwidth]{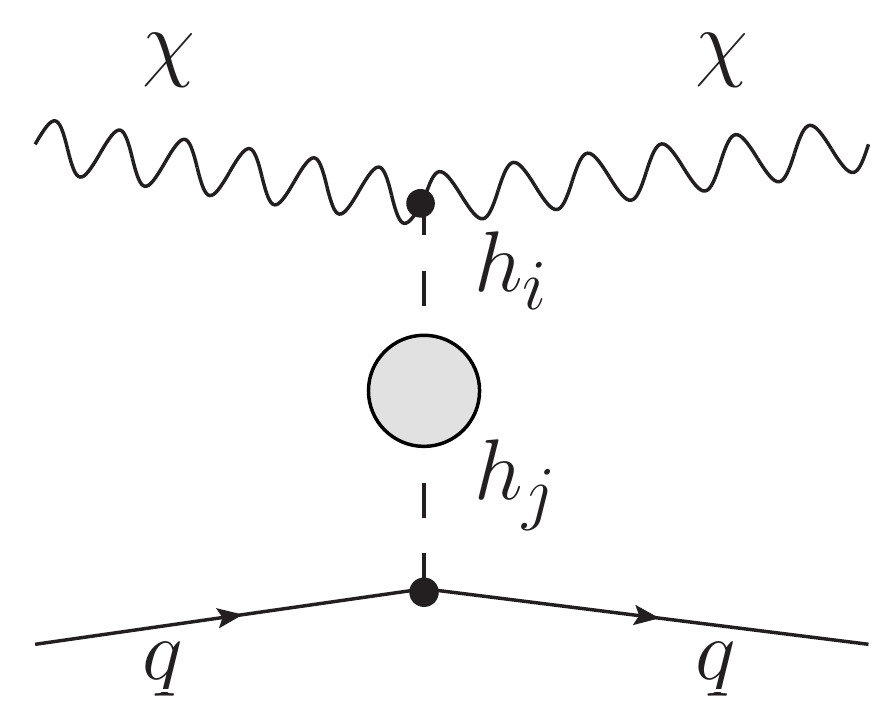}} \hspace*{0.4cm}
    \subfigure[Box Corrections\label{BOXCORRECTIONS}]{\includegraphics[width=0.18\textwidth]{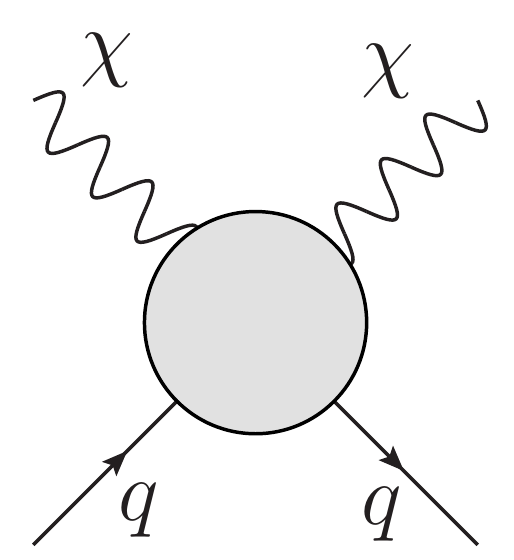}}
    \caption{Generic one-loop corrections to the scattering of VDM
      with the nucleon. The grey blob corresponds to the renormalized
      one-loop corrections. The corrections can be separated into
      vertex (a), mediator (b) and box corrections (c).}
    \label{OneLoop::Tops}
\end{figure}

As a next step, we want to include the NLO EW corrections in the
calculation of the SI cross section. For this, we evaluate the
one-loop contributions to the Wilson coefficients
  $f_q$ and $f_G$ in front of the operators in
\cref{tree::effoperators}. At this order, also the Wilson coefficient $g_q$ is
  non-zero. The additional topologies contributing
at NLO~EW are depicted in \cref{OneLoop::Tops}.  
Note that we do not include vertex corrections to the $h_i \bar q q$
vertex. They are partly given by the nuclear matrix elements and
beyond the scope of our study. For the purpose of our investigation,
we assume them to be encoded in the effective coupling factors of the
respective nuclear matrix elements. In the following, we present the
calculation of each topology separately. 

%%%%%%%%%%%%%%%%%%%%%%%%%%%%%%%%%%%%%%%%%%%%%%%%%%%%%%%%%%%%%
\subsection{Vertex Corrections $\chi\chi h_i$}
The effective one-loop coupling $\chi\chi h_i$ is extracted by
considering loop corrections to the coupling $\chi\chi
  h_i$, where we take the DM particles to be on-shell and assume a vanishing
  momentum for the Higgs boson $h_i$.
The amplitude for the NLO vertex including the
  polarisation vectors $\varepsilon^{(*)}$ of the external VDM particles, is given by 
\begin{equation}
    \ii \mathcal A_{\chi\chi h_i}^{\text{NLO}} = \ii
    \mathcal{A}_{\chi\chi h_i}^{\text{LO}}+\ii \mathcal{A}_{\chi\chi
      h_i}^{\text{VC}}+\ii \mathcal{A}_{\chi\chi h_i}^{\text{CT}}\,,
\end{equation}
with the leading-order amplitude $\ii \mathcal{A}_{\chi\chi
  h_i}^{\text{LO}}$, the virtual vertex corrections $\ii
\mathcal{A}_{\chi\chi h_i}^{\text{VC}}$ and the vertex counterterm
$\ii \mathcal{A}_{\chi\chi h_i}^{\text{CT}}$. 
Denoting by $p$ the four-momentum of the incoming VDM particle, the tree-level
amplitude is given by  
\begin{equation}
    \ii \mathcal{A}_{\chi\chi h_i}^{\text{LO}} = g_{\chi\chi h_i}
    \varepsilon(p)\cdot\varepsilon^*(p) =  2 \gX \mX  
    \varepsilon(p)\cdot\varepsilon^*(p)\begin{cases}
        \sin\alpha\,, \quad i=1\\
        \cos\alpha\,, \quad i=2
    \end{cases} \,.
\end{equation}
The vertex counterterm amplitudes for $i=1,2$ read
\begin{subequations}    
    \begin{align}
        &\ii \mathcal{A}^{\text{CT}}_{\chi\rightarrow\chi h_1} =
          \left[\frac{1}{2}\cbrak{g_{\chi\chi h_2}\delta
          Z_{h_2h_1}+g_{\chi\chi h_1}\delta Z_{h_1h_1}} + g_{\chi\chi
          h_1} \delta Z_{\chi\chi}+\delta g_{\chi\chi
          h_1}\right]\varepsilon(p)\cdot\varepsilon^*(p)\\ 
        &\ii \mathcal{A}^{\text{CT}}_{\chi\rightarrow\chi h_2} =
          \left[\frac{1}{2}\cbrak{g_{\chi\chi h_1}\delta
          Z_{h_1h_2}+g_{\chi\chi h_2}\delta Z_{h_2h_2}} + g_{\chi\chi
          h_2} \delta Z_{\chi\chi}+\delta g_{\chi\chi
          h_2}\right]\varepsilon(p)\cdot\varepsilon^*(p)\,, 
    \end{align}
    \label{XXH::CT}
\end{subequations}
with the counterterms $\delta g_{\chi\chi h_i}$ ($i=1,2$) for the couplings
\beq
g_{\chi \chi h_1} &=& 2 \gX \mX \sin \alpha \\
g_{\chi \chi h_2} &=& 2 \gX \mX \cos \alpha 
\eeq
derived from 
\begin{equation}
    \delta g_{\chi\chi h_i}  = \sum_{p} \frac{\partial g_{\chi\chi
        h_i}}{\partial p}\,,\quad p\in \{\mX, \gX,\alpha\}\,.
\end{equation}
\begin{figure}
    \centering
    \footnotesize
    \stackunder[5pt]{\includegraphics[width=0.15\textwidth]{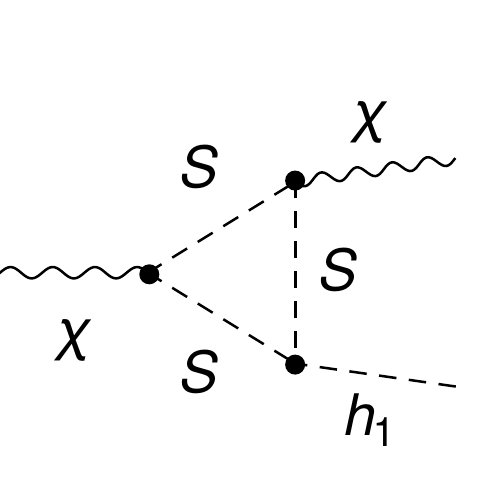}}{\tiny $S=\lbrace h_i,G_{\chi}\rbrace$}%
    \hspace{0.25cm}%
    \stackunder[5pt]{\includegraphics[width=0.15\textwidth]{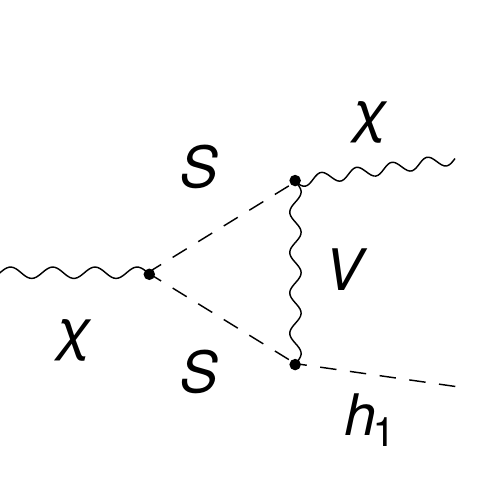}}{\tiny $S,V=\lbrace h_i,G_{\chi}\rbrace,\lbrace X \rbrace$}%
    \hspace{0.25cm}%
    \stackunder[5pt]{\includegraphics[width=0.15\textwidth]{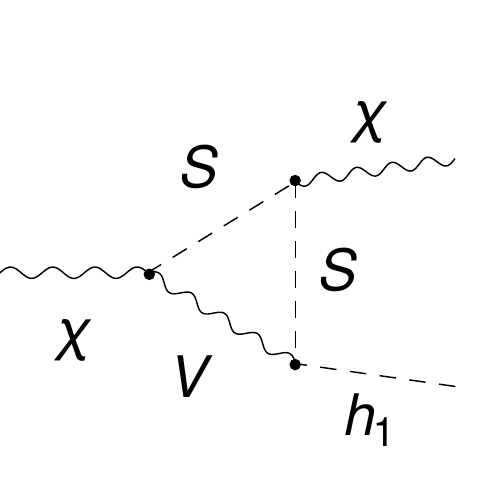}}{\tiny $S,V=\lbrace h_i,G_{\chi}\rbrace,\lbrace X \rbrace$}%
    \hspace{0.25cm}%
    \stackunder[5pt]{\includegraphics[width=0.15\textwidth]{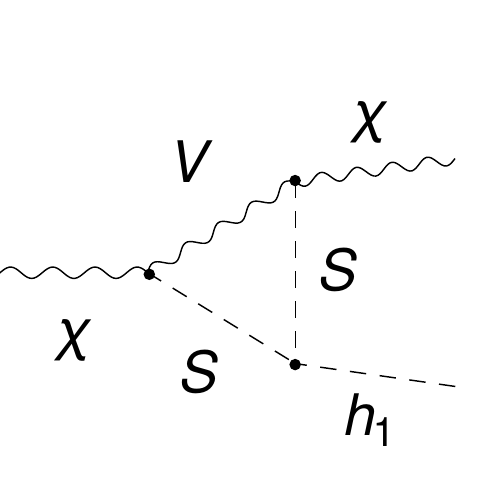}}{\tiny $S,V=\lbrace h_i\rbrace,\lbrace X \rbrace$}%
    \hspace{0.25cm}\\%
    \stackunder[5pt]{\includegraphics[width=0.15\textwidth]{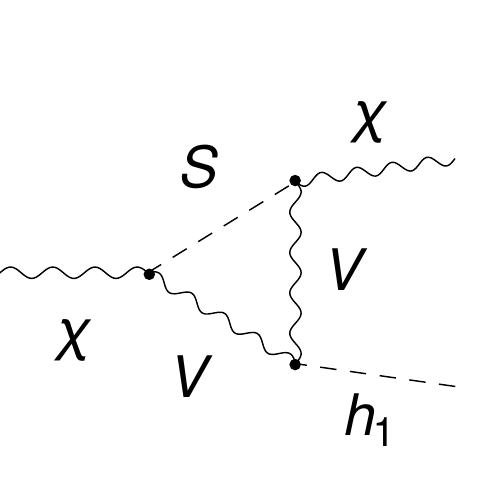}}{\tiny $S,V=\lbrace h_i\rbrace,\lbrace X \rbrace$}%
    \hspace{0.25cm}%
    \stackunder[5pt]{\includegraphics[width=0.15\textwidth]{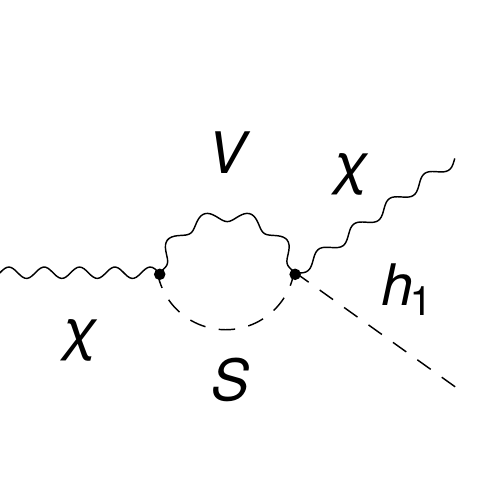}}{\tiny $S,V=\lbrace h_i\rbrace,\lbrace X \rbrace$}%
    \hspace{0.25cm}%
    \stackunder[5pt]{\includegraphics[width=0.15\textwidth]{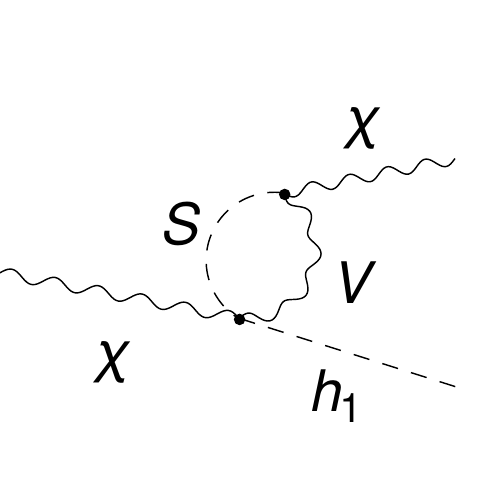}}{\tiny $S,V=\lbrace h_i\rbrace,\lbrace X \rbrace$}%
    \hspace{0.25cm}%
    \stackunder[5pt]{\includegraphics[width=0.15\textwidth]{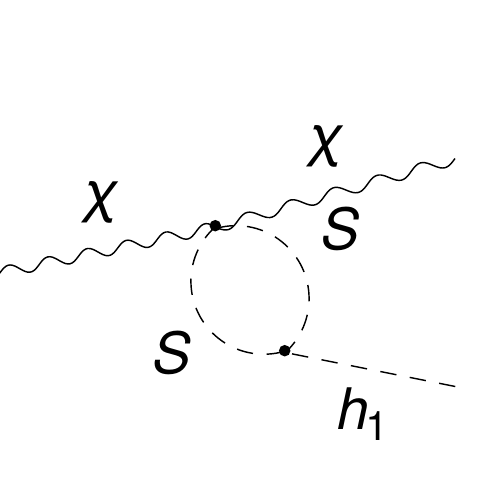}}{\tiny $S=\lbrace h_i , G_{\chi}\rbrace$}%
    \hspace{0.25cm}%
\caption{Generic diagrams contributing to the virtual corrections
      to the vertex $\chi\chi h_i$. The generic symbols denote $F$
      fermions, $S$ scalars and $V$ gauge bosons.}
    \label{XXh::VC}
\end{figure}
In \cref{XXh::VC} all contributing NLO diagrams are shown, where $S$
denotes scalars, $F$ fermions and $V$ vector bosons. At NLO an
additional tensor structure arises in the amplitude. Let $p_{\text{in}}$ be
the incoming momentum  and $p_{\text{out}}$ the outgoing momentum of the DM
vector gauge boson. Assuming zero momentum transfer is equivalent to assuming
$p_{\text{in}}=p_{\text{out}}$. Note that this assumption is stricter than simply
assuming $p_{\text{in}}^2=p_{\text{out}}^2$, since this only implies the same masses
for the incoming and outgoing particles. The additional new tensor
structure (denoted by $\sim$ NLO) is given by  
\begin{equation}
    \ii \mathcal A^{\text{NLO}} =
    \cbrak{\dots}\underbrace{\varepsilon(p_{\text{in}})\cdot\varepsilon^*(p_{\text{out}})}_{\sim
      \text{LO}} + \cbrak{\dots}\underbrace{ \cbrak{p_{\text{in}}\cdot
        \varepsilon^*(p_{\text{out}})}\cbrak{p_{\text{out}}\cdot
        \varepsilon(p_{\text{in}})}}_{\sim \text{NLO}}\,. 
\end{equation}
The additional NLO tensor structure vanishes by assuming $p_{\text{in}} =
p_{\text{out}}$, and because for freely propagating gauge bosons we
have $\varepsilon(p)\cdot p =0$. 
The counterterms in \cref{XXH::CT} cancel all UV-poles of the virtual
vertex corrections in \cref{XXh::VC} which has been checked both analytically and
numerically. Accounting for the symmetry factor of the amplitude and
projecting onto the corresponding tensor structure, the vertex corrections
are plugged in the generic diagram in \cref{VERTEXCORRECTIONS} which
contributes to the operator $\chi_{\mu}\chi^{\mu}m_q \bar q q$. We will
refer to the resulting contribution as $f_q^{\text{vertex}}$. 
Since the expression it quite lengthy, we do not give the explicit formula here.

%%%%%%%%%%%%%%%%%%%%%%%%%%%%%%%%%%%%%%%%%%%%%%%%%%%%%%%%%%%%%
\subsection{Mediator Corrections}
We proceed in a similar way for the mediator corrections. We calculate
the self-energy corrections to the two-point functions with all
possible combinations of external Higgs fields and plug
these into the one-loop propagator in the generic amplitude in
\cref{MEDIATORCORRECTIONS}. The self-energy contribution to the $h_i h_j$
propagator ($i,j=1,2$) reads  
\begin{equation}
    \Delta_{h_ih_j} =
    -\frac{\hat{\Sigma}_{h_ih_j}(p^2=0)}{m_{h_i}^2m_{h_j}^2} \;,
\end{equation}
with the renormalised self-energy matrix
\begin{equation}
  \begin{pmatrix}
      \hat{\Sigma}_{h_1h_1}&\hat{\Sigma}_{h_1h_2}\\
      \hat{\Sigma}_{h_2h_1}&\hat{\Sigma}_{h_2h_2}
  \end{pmatrix} 
  \equiv
   \hat\Sigma(p^2) = \Sigma(p^2) - \delta m^2 -\delta T +\frac{\delta Z}{2} \cbrak{p^2-\mathcal{M}^2} + \cbrak{p^2-\mathcal{M}^2}\frac{\delta Z}{2}\,,
\end{equation}
where the mass matrix $\mathcal{M}$ and the tadpole counterterm matrix
$\delta T$ are defined in \cref{RENORM::MASSMATRIX}. The $Z$-factor
matrix $\delta Z$ corresponds to the matrix with the components
$\delta Z_{h_ih_j}$ defined in \cref{RENORM::ZFACTOR}.  
Projecting the resulting one-loop correction on the corresponding
tensor structure, we obtain the effective one-loop correction to the
Wilson coefficient of the operator $\chi_{\mu}\chi^{\mu}m_q \bar q q$ induced by the mediator
corrections as 
\begin{equation}
    f_q^{\text{med}} = \frac{g \gX \mX}{2m_W} \sum_{i,j} R_{\alpha,i2}R_{\alpha,j1}\Delta_{h_ih_j}\,,
\end{equation} 
with the rotation matrix $R_{\alpha}$ defined in \cref{VDM::massbasis}.

%%%%%%%%%%%%%%%%%%%%%%%%%%%%%%%%%%%%%%%%%%%%%%%%%%%%%%%%%%%%%
\subsection{Box Corrections \label{sec:boxcorrections}}
\begin{figure}
    \centering
    \footnotesize
    \stackunder[5pt]{\includegraphics[width=0.15\textwidth]{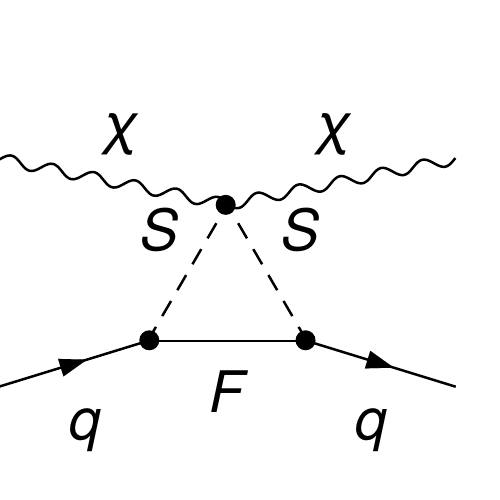}}{\tiny $F,S=\lbrace q\rbrace,\lbrace h_i\rbrace$}%
    \hspace{0.25cm}%
    \stackunder[5pt]{\includegraphics[width=0.15\textwidth]{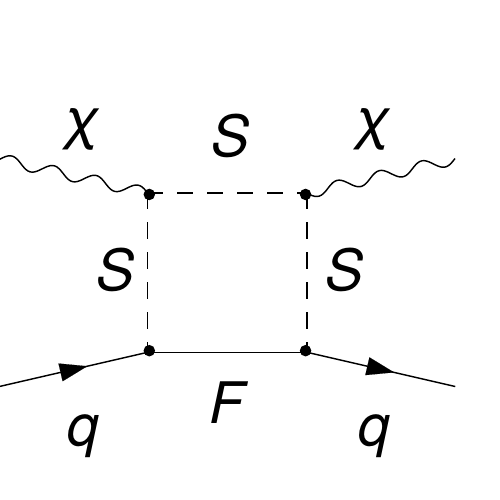}}{\tiny $F,S=\lbrace q\rbrace,\lbrace h_i,G_{\chi}\rbrace$}%
    \hspace{0.25cm}%
    \stackunder[5pt]{\includegraphics[width=0.15\textwidth]{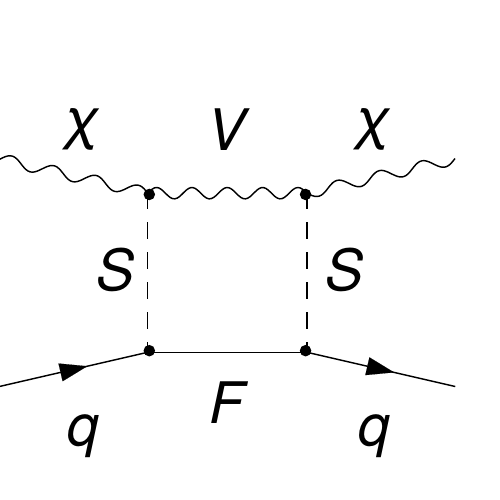}}{\tiny $F,S,V=\lbrace q\rbrace,\lbrace h_i,G_{\chi}\rbrace,\lbrace X \rbrace$}%
    \hspace{0.25cm}%
    \stackunder[5pt]{\includegraphics[width=0.15\textwidth]{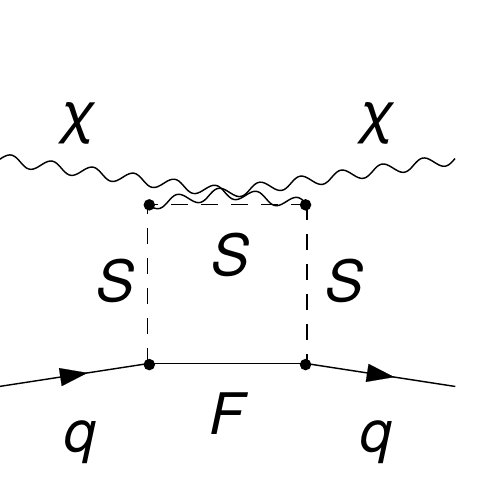}}{\tiny $F,S=\lbrace q\rbrace,\lbrace h_i\rbrace$}%
    \hspace{0.25cm}%
    \stackunder[5pt]{\includegraphics[width=0.15\textwidth]{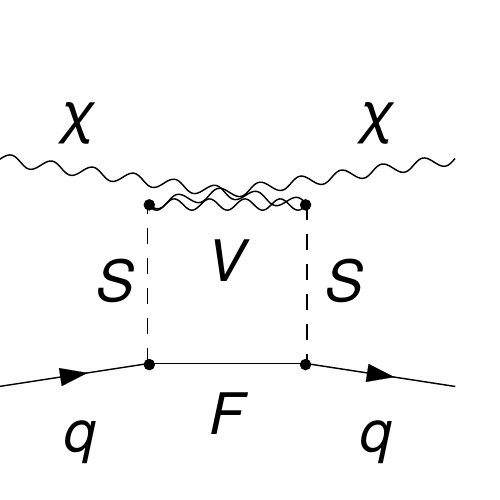}}{\tiny $F,S,V=\lbrace q\rbrace,\lbrace h_i,G_{\chi}\rbrace,\lbrace X \rbrace$}%
    \caption{Generic diagrams of the box topology contributing to the
      SI cross section. The symbol $S$ denotes scalars, $F$
      fermions and $V$ vector bosons. The flavour of the fermion $F$
      and the external quark $q$ are the same as we set the CKM matrix
    equal to the unit matrix. \label{fig:genbox}} 
\end{figure}
We now turn to the box corrections. The generic set of diagrams representative of the box topology is depicted in Fig.~\ref{fig:genbox}. In the following, we present the treatment of box diagrams contributing to the SI cross section. In order to extract for the
spin-independent cross section the relevant tensor structures from the
box diagram, we expand the loop diagrams in terms of the momenta $p_q$
of the external quark that is not relativistic \cite{Abe:2018emu}.
Since we are considering zero momentum transfer, the incoming and
outgoing momenta of the quark are 
the same, 
\begin{equation}
    p_q^{\text{in}} = p_q^{\text{out}}\,.
    \label{OneLoop::Momenta}
\end{equation}
Note that as in the case of the vertex corrections this requirement is
stricter than requiring that the squared 
momenta are the same, since this only implies same masses for incoming
and outgoing particles. Assuming small quark momenta, and because
the mass of the light quarks is much smaller than the energy scale of
the interaction, allows for the 
simplification of the propagator terms arising in the box diagrams
through the expansion, 
\begin{equation}
    \frac{1}{(l\pm p_q)^2-m_q^2} = \frac{1}{l^2} \mp \frac{2 p_q \cdot
      l}{l^4}+\mathcal O (p_q^2 /l^4) \,,
\end{equation}
where $l$ is the loop momentum of the box diagram, $m_q$ the mass of
the quark and where we use $m_q^2 = p_q^2$. After applying this
expansion to the box diagrams, the result has to be projected onto the
required tensor structures contributing to the operators in
\cref{tree::effoperators}. The box diagrams contribute to 
$ X_{\mu} X^{\mu} m_q \bar qq $ and the twist-2 operators.  
By rewriting\cite{Ertas:2019dew,Hisano:2010ct,Hisano:2015bma}
\begin{equation}
    \bar q\ii \partial_{\mu}\gamma_{\nu}q = \mathcal{O}_{\mu\nu}^q + \bar q\frac{\ii \partial_{\mu}\gamma_{\nu}-\ii \partial_{\nu}\gamma_{\mu}}{2} q + \frac{1}{4} g_{\mu\nu} m_q \bar qq\,,
    \label{BOX::rewrite}
\end{equation}
the parts of the loop amplitude that correspond to the twist-2 and the $
X_{\mu} X^{\mu} m_q \bar qq $ operator can be extracted. The asymmetric part in 
\cref{BOX::rewrite} does not contribute to the SI cross section and
therefore can be dropped. We refer to these one-loop contributions to
the corresponding tree-level Wilson coefficients as $f_q^{\text{box}}$
and $g_q^{\text{box}}$. \s

\begin{figure}
    \centering
\hspace*{-4.5cm}
    \subfigure{\includegraphics[width = 0.20\textwidth]{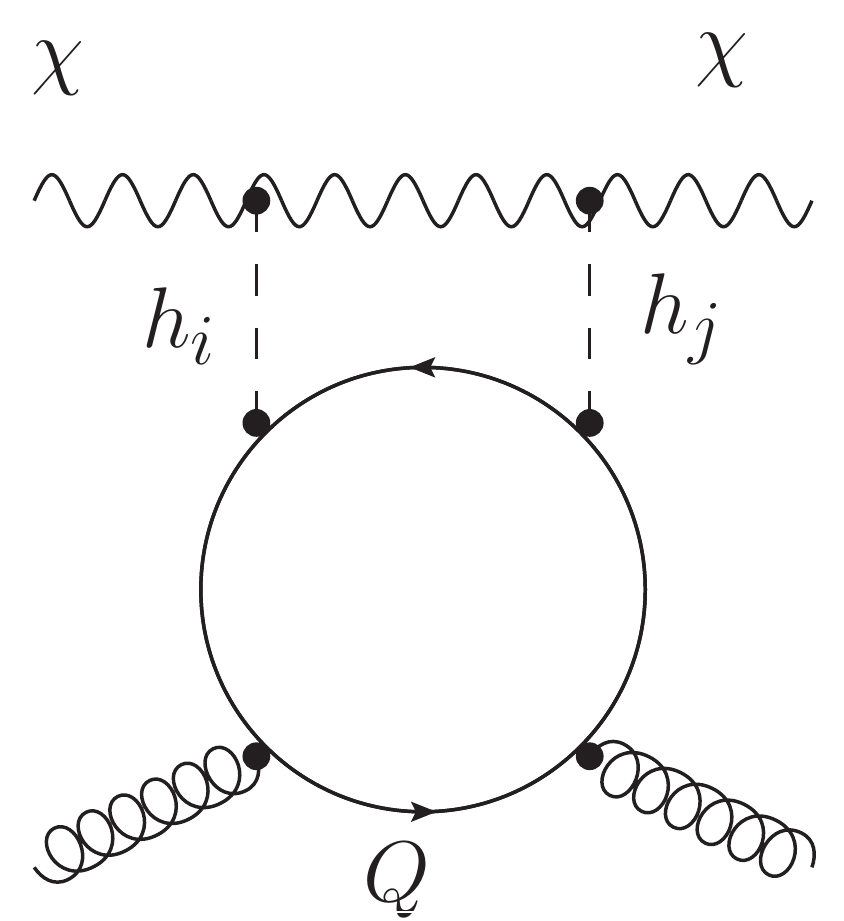}}
\\[-3cm]
\hspace*{6cm}
    \subfigure{\includegraphics[width =
      0.25\textwidth]{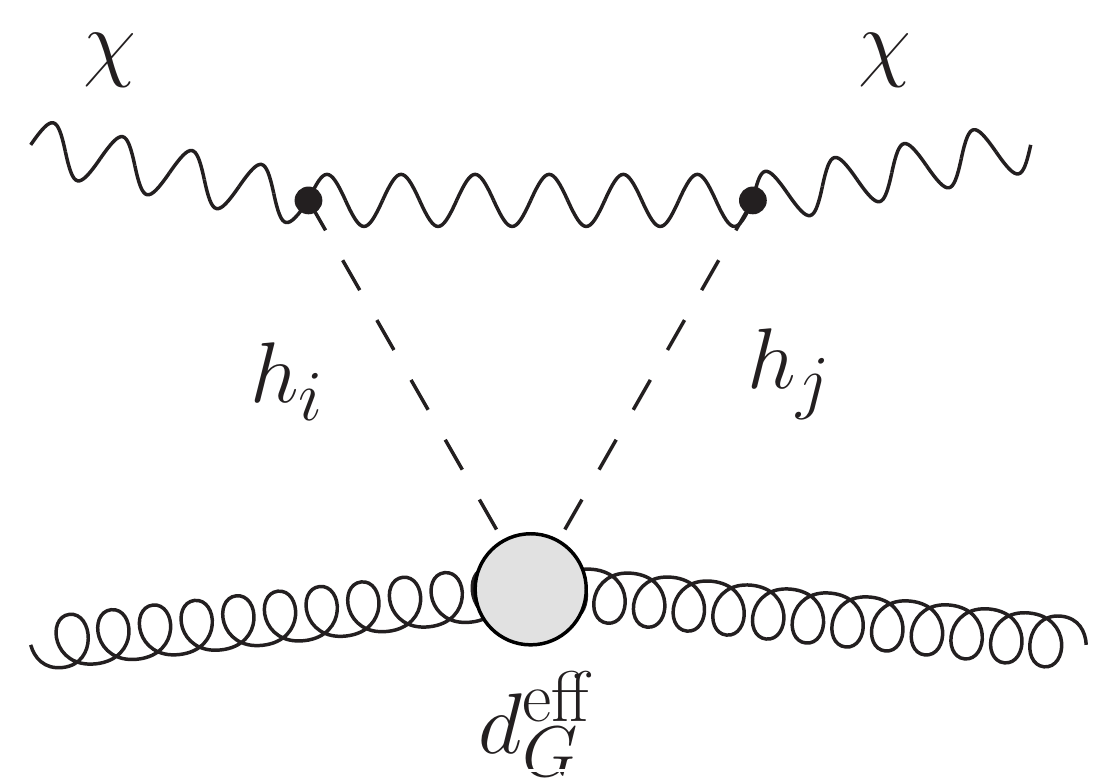}}
\vspace*{0.3cm}
    \caption{The full two-loop gluon interaction with the DM candidate
      (left) and the effective two-loop interaction after integration
      out the heavy quarks (right). \label{BOX::GLU}} 
\end{figure}

As discussed in Refs.~\cite{Ertas:2019dew,Abe:2018emu} the box
diagrams also induce additional contributions to the effective gluon
interaction with the VDM particle that have to be taken into account
in the Wilson coefficient $f_G$ in Eq.~(\ref{tree::effoperators}b). The
naive approach of using the same replacement as in \cref{tree::gluonmap}
to obtain the gluon interaction induces large
  errors \cite{Abe:2018emu}. To circumvent the over-estimation
of the gluon interaction without performing the full two-loop
calculation, we adopt the ansatz proposed in Ref.~\cite{Ertas:2019dew}. For
heavy quarks compared to the mediator mass, it is possible to
derive an effective coupling between
two Higgs bosons and the gluon fields. Using the Fock-Schwinger gauge allows us to express the gluon fields in terms of the field
strength tensor $G^a_{\mu\nu}$, simplifying the extraction of the
effective two-loop contribution to $f_G$. Integrating out the
top-quark yields the following effective two-Higgs-two-gluon
coupling\cite{Ertas:2019dew}\footnote{The authors of
    Ref.~\cite{Ertas:2019dew} found that the bottom and charm quark
    contributions are small. This may not be the case if the Higgs couplings
    to down-type quarks are enhanced. This does not apply for our
    model, however.}
\begin{equation}
    \mathcal L^{hhGG} = \frac{1}{2} d_G^{\text{eff}} h_i h_j
    \frac{\alpha_S}{12\pi} G^a_{\mu\nu}G^{a \,\mu\nu}\,, 
    \label{box::dgeff}
\end{equation}
where the effective coupling $d_G^{\text{eff}}$ of
Ref.~\cite{Ertas:2019dew} has to be adopted to our model. First of all
we only have scalar-type mediators, given by the Higgs bosons $h_i$, so that the
mixing angle $\phi_{\text{SM}}$ of Ref.~\cite{Ertas:2019dew} which
quantifies the CP-odd admixture, is set to  
\begin{equation}
    \phi_{\text{SM}} = 0\,.
\end{equation}
Second, the coupling of the Higgs bosons $h_i$ to the top quark
differs depending on which Higgs boson is coupled, so that the
effective coupling in \cref{box::dgeff} becomes 
\begin{equation}
    d_G^{\text{eff}} \to \cbrak{d_G^{\text{eff}}}_{ij} =
    (R_\alpha)_{i1} (R_\alpha)_{j1} \frac{1}{v^2}\,,
\end{equation}
with the rotation matrix $R_\alpha$ defined in \cref{VDM::massbasis}. 
The effective coupling allows for the calculation of the box-type
diagram in \cref{BOX::GLU} (right). \s

In Ref.~\cite{Ertas:2019dew}, the full two-loop calculation was
performed. The comparison with the complete two-loop result showed
very good agreement between the approximate and the exact result for
mediator masses below $m_t$. Our model is structurally not different
in the sense that the mediator coupling to the DM particle (a fermion
in Ref.~\cite{Ertas:2019dew}) is also a scalar particle so that the results
obtained in Ref.~\cite{Ertas:2019dew} should be applicable to our model as
well. Moreover, the box contribution to the NLO SI direct detection
cross section is only minor as we verified explicitly. \s

The diagram in \cref{BOX::GLU} (right) yields the following contribution to the
Lagrangian 
\begin{equation}
  \mathcal{L}_{\text{eff}}\supset  
\left( d_G^{\text{eff}}\right)_{ij} C^{ij}_\triangle \chi_{\mu}\chi^{\mu} \frac{-\alpha_S}{12\pi}
  G^a_{\mu\nu}G^{a \, \mu\nu}\,, 
\end{equation} 
where $C^{ij}_\triangle$ denotes the contribution from the triangle
loop built up by $h_i$, $h_j$ and the VDM particle. It has to be
extracted from the calculated amplitude of 
\cref{BOX::GLU} (right). Using \cref{eq:ggxx}  the contributions by
the box topology to the gluon-DM interaction are given by
\begin{equation}
    f_G^{\text{top}}= \left( d_G^{\text{eff}} \right)_{ij} C^{ij}_\triangle \frac{-\alpha_S}{12 \pi} \,.
\end{equation}

%%%%%%%%%%%%%%%%%%%%%%%%%%%%%%%%%%%%%%%%%%%%%%%%%%%%%%%%%%%%
\subsection{The SI One-Loop Cross Section}
In the last sections we discussed the extraction of the one-loop
effective form factors for the operators in
\cref{tree::effoperators}. The NLO EW SI cross section can then be
obtained by using the effective one-loop form factor 
\begin{equation}
    \frac{f_N^{\text{NLO}}}{m_N} = \sum_{q=u,d,s} f_q^{\text{NLO}}
    f^N_{T_q} + \sum_{q=u,d,s,c,b}\frac{3}{4} \cbrak{q(2)+\bar q(2)}
    g_q^{\text{NLO}} - \frac{8\pi}{9\alpha_S} f^N_{T_G}
    f_G^{\text{NLO}}\,, 
\label{eq:sigmanlo}
\end{equation}
with the Wilson coefficients at one-loop level given by
\begin{subequations}
    \begin{align}
    &f_q^{\text{NLO}} =
      f_q^{\text{vertex}}+f_q^{\text{med}}+f_q^{\text{box}}
      \\
    &g_q^{\text{NLO}} = g_q^{box} \\
    &f_G^{\text{NLO}} = -\frac{\alpha_S}{12\pi}
      \sum_{q=c,b,t}\cbrak{f_q^{\text{vertex}}+f_q^{\text{med}}} +
      f_G^{\text{top}}\,. 
%   &f_q^{\text{NLO}} =
%      f_q^{\text{LO}}+f_q^{\text{vertex}}+f_q^{\text{med}}+f_q^{\text{box}}
%      \\
%    &g_q^{\text{NLO}} = g_q^{box} \\
%    &f_G^{\text{NLO}} = -\frac{\alpha_S}{12\pi}
%      \sum_{q=c,b,t}\cbrak{f_q^{\text{LO}}+f_q^{\text{vertex}}+f_q^{\text{med}}} +
%      f_G^{\text{top}}\,. 
\end{align}
\end{subequations}
Like at LO, the heavy quark contributions of
  $f_q^{\text{vertex}}$ and $f_q^{\text{med}}$ have to be attributed
  to the effective gluon interaction. With the LO form factor given by
\beq
\frac{f_N^{\text{LO}}}{m_N} = f_q^{\text{LO}} \left[\sum_{q=u,d,s} f^N_{T_q}
    + \sum_{q=c,b,t} \frac{2}{27} f_{T_G}^N \right] \,,
\eeq
where $f_q^{\text{LO}}$ has been given in
Eq.~(\ref{eq:fqlo}), we have for the NLO EW SI cross section at leading
order in $\alpha_S$,
\beq
\sigma_N = \frac{1}{\pi} \left( \frac{m_N}{\mX + m_N}\right)^2
\left[ |f_N^{\text{LO}}|^2 + 2 \mbox{Re} \left( f_N^{\text{LO}}
  f_N^{\text{NLO}*} \right) \right]\;.
\eeq

%%%%%%%%%%%%%%%%%%%%%%%%%%%%%%%%%%%%%%%%%%%%%%%%%%%%%%%%%%%%%
\section{Numerical Analysis \label{sec:result}}
In our numerical analysis we use parameter points
that are compatible with current theoretical and experimental
constraints. These are obtained by performing a scan in the parameter
space of the model and by checking each data set for compatibility
with the constraints. In order to do so, the VDM model was implemented
in the code \texttt{ScannerS}~\cite{Coimbra:2013qq, Costa:2015llh}
which automatises the parameter scan. We require the SM-like
Higgs boson (denoted by $h$ in the following) to have a mass of
$m_h=125.09$~GeV~\cite{Aad:2015zhl}. With {\tt ScannerS}, we check if
the minimum of the potential is the global one and if the generated points satisfy the
theoretical constraints of boundedness from below and perturbative
unitarity. We furthermore impose the perturbativity constraint $g_\chi^2
< 4 \pi$. Furthermore, the model has to comply with the experimental Higgs data. 
In the VDM model, the Higgs couplings to the SM particles
are modified by a common factor given in terms of the mixing angle
$\alpha$, that is hence constrained by the combined values for the
signal strengths~\cite{Aad:2015zhl}. Through an interface with
\texttt{HiggsBounds}~\cite{Bechtle:2008jh,Bechtle:2011sb,Bechtle:2013wla}
we additionally check for collider bounds from LEP, Tevatron and
the LHC. We require agreement with the exclusion limits derived for
the non-SM-like Higgs boson at 95\% confidence level. 
Among these searches the most stringent bound arises from the search
for heavy $ZZ$ resonances~\cite{Aaboud:2017rel}. Still, the bounds for
the mixing angle $\alpha$ derived from the measurement of the Higgs
couplings are by far the most relevant. 
In order to check for the constraints from the Higgs data, the Higgs
decay widths and branching ratios were calculated with {\tt
  sHDECAY}~\cite{Costa:2015llh}\footnote{The program {\tt sHDECAY} can
  be downloaded from the url:
  \url{http://www.itp.kit.edu/~maggie/sHDECAY}.}, which includes the
state-of-the-art higher-order QCD corrections. The code {\tt sHDECAY}
is based on the implementation of the models in {\tt
  HDECAY}~\cite{Djouadi:1997yw,Djouadi:2018xqq}. \s

Concerning the DM constraints, information on the DM particle from
LHC searches through the invisible width of the SM Higgs boson were
taken into account~\cite{Bechtle:2008jh,Bechtle:2011sb,Bechtle:2013wla}. 
Furthermore, the DM relic abundance has been calculated with
\texttt{MicrOMEGAs}~\cite{Belanger:2006is,Belanger:2007zz,Belanger:2010pz,Belanger:2013oya},
and compared with the current experimental result from the Planck
Collaboration~\cite{Ade:2015xua}, 
\beq
({\Omega}h^2)^{\rm obs}_{\rm DM} = 0.1186 \pm 0.002 \;. 
\eeq
We do not force the DM relic abundance to be in
this interval, but rather require the calculated abundance to be equal
to or smaller than the observed one. Hence, we allow the DM to not
saturate the relic density and therefore define a DM fraction 
\begin{eqnarray}
f_{\chi\chi} = \frac{({\Omega} h^2)_{\chi}}{(\Omega h^2)^{\rm
  obs}_{\text{DM}}}\,, 
\label{eq:dmfraction}
\end{eqnarray} 
where $(\Omega h^2)_{\chi}$ stands for the calculated DM relic
abundance of the VDM model. 
DM indirect detection also provides constraints on the VDM model. The
annihilation into visible states, mainly  into $ZZ$, $W^+ W^-$,
$b\bar{b}$ and light quark pairs, can be measured by
Planck~\cite{Ade:2015xua}, if it manifests itself in anisotropies of
the cosmic microwave background (CMB), by
Fermi-LAT~\cite{Ackermann:2015zua} if it comes form the $\gamma$-ray
signals in the spheroidal dwarf galaxies, and by
AMS-02~\cite{Aguilar:2014fea,Accardo:2014lma} if it originates from
$e^\pm$ excesses in the Milky Way. As shown in  Ref.~\cite{Elor:2015bho},
the Fermi-LAT upper bound on the DM annihilations is the most
stringent one. In order to obtain the bound, we
follow Ref.~\cite{Ackermann:2015zua} in their claim that all final states
give approximately the same upper bound on the DM annihilation cross
sections. Hence we use the Fermi-LAT bound
from Ref.~\cite{Ackermann:2015zua} on $b\bar{b}$ when $\mX
\geqslant m_b$, and on light quarks for $\mX<
m_b$. In the comparison with the data, the DM fraction in
Eq.~(\ref{eq:dmfraction}) has to be taken into account, and an
effective DM annihilation cross section is defined by 
\beq
\sigma^{\rm eff}_{\chi\chi} = f^2_{\chi\chi} \sigma_{\chi\chi} \;,
\label{eq:sigmaeffannihil}
\eeq 
with $f_{\chi\chi}$ and $\sigma_{\chi\chi}$ computed by \texttt{MicrOMEGAs}. 
\s

The sample was generated taking into account the experimental bounds
on the DM nucleon SI cross section at LO. The most stringent bound on
this cross section is the one from the
XENON1T~\cite{Aprile:2017iyp,Aprile:2018dbl} experiment. We apply the
latest XENON1T upper bounds~\cite{Aprile:2018dbl} for a DM mass above
6~GeV and the combined limits from CRESST-II~\cite{Angloher:2015ewa} and 
 CDMSlite~\cite{Agnese:2015nto} are used for lighter DM particles. 
Note that the experimental limits on DM-nucleon scattering were derived by
  assuming that the DM candidate makes up for all of the DM
  abundance. Hence, the correct quantity to be directly compared
  with experimental limits is the effective DM-nucleon
  cross-section defined by
\beq 
\sigma^{\rm eff}_{\chi N} \equiv f_{\chi\chi} \sigma_{\chi N} \;,
\label{cor_factor}
\eeq
where $\chi N$ stands for the scattering VDM $\chi$ with the
nucleon $N$, and $f_{\chi\chi}$ denotes the respective DM fraction,
defined in Eq.~(\ref{eq:dmfraction}). The formula for the LO direct
detection cross section $\sigma_{\chi N}$ in our VDM model has been
given in Eq.~(\ref{eq:siddlo}) and the NLO contributions have been
discussed in Section~\ref{sec:dd1loop}. For our numerical analysis, we
use the LO and NLO results for $N=p$. \s

The ranges of the input parameters of the scan performed to generate
viable parameter sets are listed in Table~\ref{tab:vdmscan}. From
here on, we denote the non-SM-like of the two CP-even Higgs bosons
$m_{h_i}$ ($i=1,2$) by $m_\phi$, the SM-like Higgs boson is
called $m_h$. Note, that in {\tt ScannerS} the scan is performed over
$\mX$ and $v_S$ instead of $\mX$ and $\gX$. The corresponding $\gX$
values are given by $\gX=\mX/v_S$. Only points with $\gX^2 \le 4 \pi$
are retained. Note that we vary $\alpha$ in the range $[-\pi/4,
\pi/4]$ to optimize the scan. This is possible due to the bound on
$\sin \alpha$ that comes from the combined signal strength
measurements of the production and decay of the SM-like Higgs
boson~\cite{Aad:2015zhl}. 
\begin{table}
\begin{center}
\begin{tabular}{l|cccc} \toprule
& $m_{\phi}$ [GeV] & $\mX$ [GeV] & $v_S$ [GeV] & $\alpha$ \\ \hline
%& & & & \multicolumn{11}{c}{in GeV} \\ \midrule
min & 1 & 1 & 1 & $-\frac{\pi}{4}$ \\
max & 1000 & 1000 & $10^7$ & $\frac{\pi}{4}$ \\ \bottomrule
\end{tabular}
\caption{Input parameters for the VDM model scan, all parameters
  varied independently between the given minimum and maximum
  values. The SM-like Higgs boson mass is set
  $m_h=125.09$~GeV and the SM VEV $v=246.22$~GeV. \label{tab:vdmscan}}
\end{center}
\end{table}
The remaining input parameters, gauge, lepton and quark masses,
electric coupling, Weinberg angle and weak $SU(2)$ coupling, are set to
\beq
\begin{array}{lllllllll}
m_W &=& 80.398 \mbox{ GeV } \;, \quad & m_Z &=& 91.1876 \mbox{ GeV }
\;, \quad & \sin\theta_W &=& 0.4719 \;,\\
m_e &=& 0.511\cdot 10^{-3} \mbox{ GeV } \;, \quad & m_\mu &=& 0.1057 \mbox{
GeV } \;, \quad & m_\tau &=& 1.777 \mbox{ GeV } \;, \\
m_u &=& 0.19 \mbox{ MeV } \;, \quad & m_d &=& 0.19 \mbox{ MeV } \;, \quad 
& m_s &=& 0.19 \mbox{ MeV } \;, \\
m_c &=& 1.4 \mbox{ GeV } \;, \quad & m_b &=& 4.75 \mbox{ GeV } \;,
\quad & m_t &=& 172.5 \mbox{ GeV } \;. 
%e &=& 0.30815 \;, \quad & \sin\theta_W &=& 0.4719 \;, \quad &
%g &=& 0.6531 \;.
\end{array}
\eeq
For the proton mass we take \beq
m_p = 0.93827 \mbox{ GeV}\,.
\eeq

%%%%%%%%%%%%%%%%%%%%%%%%%%%%%%%%%%%%%%%%%%%%%%%%%%%%%%%%%%%%%
\subsection{Results}
In the following we present the LO and NLO results for the
spin-independent direct detection cross section of the VDM model. We
investigate the size of the NLO corrections and their phenomenological
impact. We furthermore discuss the gauge dependence of the results and
the influence of the renormalisation scheme on the NLO results. If not
stated otherwise, results are presented for the Feynman gauge, {\it
  i.e.}~the gauge parameter $\xi$\footnote{We commonly denote by $\xi$
the gauge parameter for all gauge bosons.} is set equal to one,
$\xi=1$. In the NLO results, the default renormalisation scheme for
the mixing angle $\alpha$ is the KOSY scheme, {\it
  cf.}~Subsection~\ref{sec:alpharenorm}.

%%%%%%%%%%%%%%%%%%%%%%%%%%%%%%%%%%%%%%%%%%%%%%%%%%%%%%%%%%%%%
\subsubsection{The SI Direct Detection Cross Section at Leading Order}
\begin{figure}[ht!]
    \centering
    \includegraphics[width=0.5\textwidth]{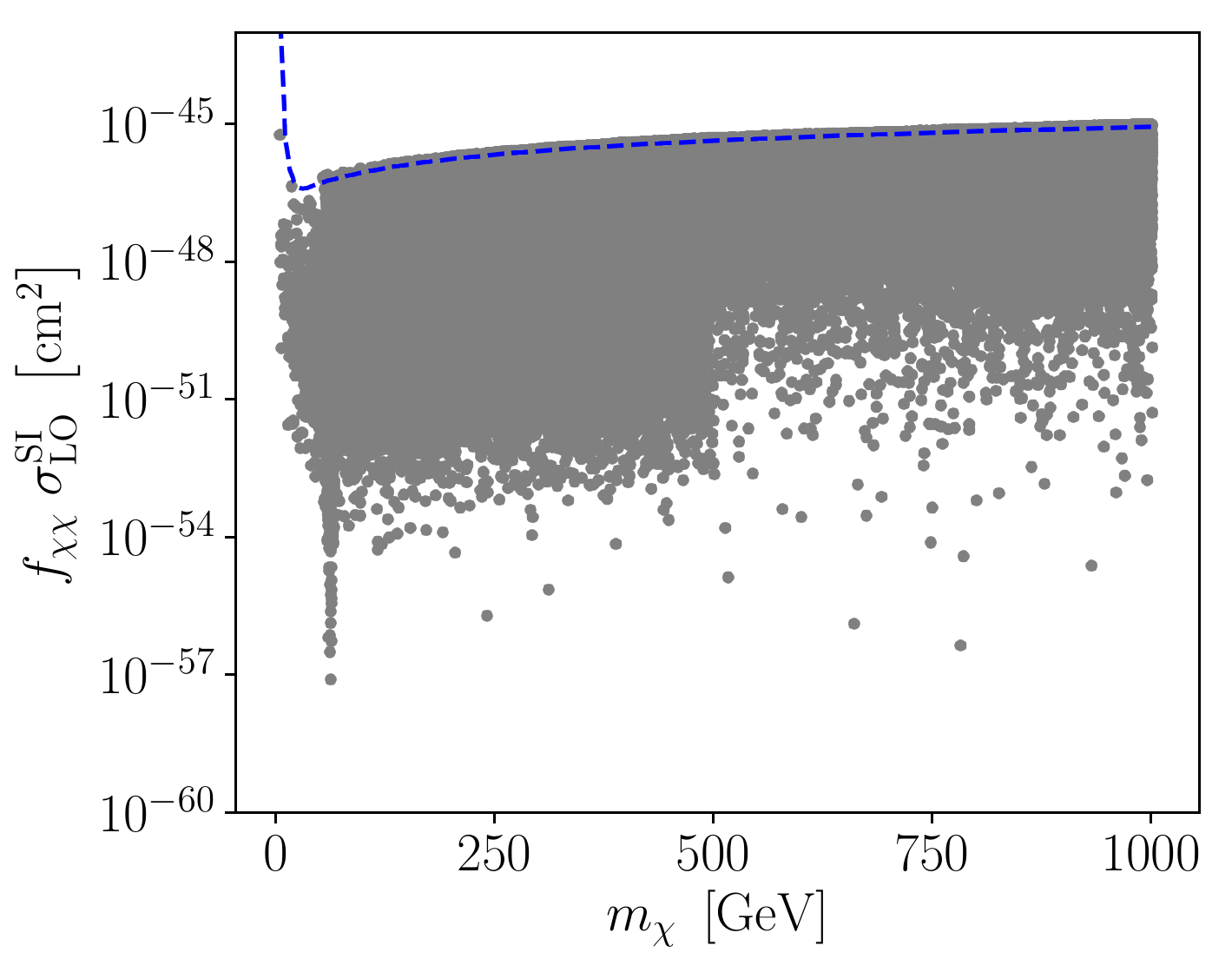}
    \caption{Grey: The tree-level SI cross section $\sigma^{\text{LO}}$ versus the
      DM mass $\mX$ in GeV for the complete parameter sample
      compatible with the applied constraints. The blue line denotes the
      Xenon Limit.}
    \label{result1}
\end{figure}

In Fig.~\ref{result1} we show in grey the LO results of the direct detection
cross section for all points of the VDM model that are compatible with
our applied constraints, as a function of the DM mass $m_\chi$. Note,
that we also include the perturbativity limit on $\gX$,
$\gX^2<4\pi$. The result is compared to the
Xenon limit shown in blue. Note that, in order to be able to
  compare with the Xenon limit, we applied the correction factor
  $f_{\chi\chi}$ to the LO and NLO direct detection cross section, {\it
    cf.}~Eq.~(\ref{cor_factor}). Since the compatibility with the Xenon
limit is already included in the selection of valid parameter points,
all cross section values lie below the blue line (modulo the size of
the grey points). As can be inferred from Fig.~\ref{result1}, the LO cross section can be
substantially smaller than the present sensitivity of the Xenon
experiment, by more than 10 orders of magnitude.  

%%%%%%%%%%%%%%%%%%%%%%%%%%%%%%%%%%%%%%%%%%%%%%%%%%%%%%%%%%%%
\subsubsection{Results for $m_\phi < m_t$}
\begin{figure}[ht!]
    \centering
    \subfigure{\includegraphics[width=0.45\textwidth]{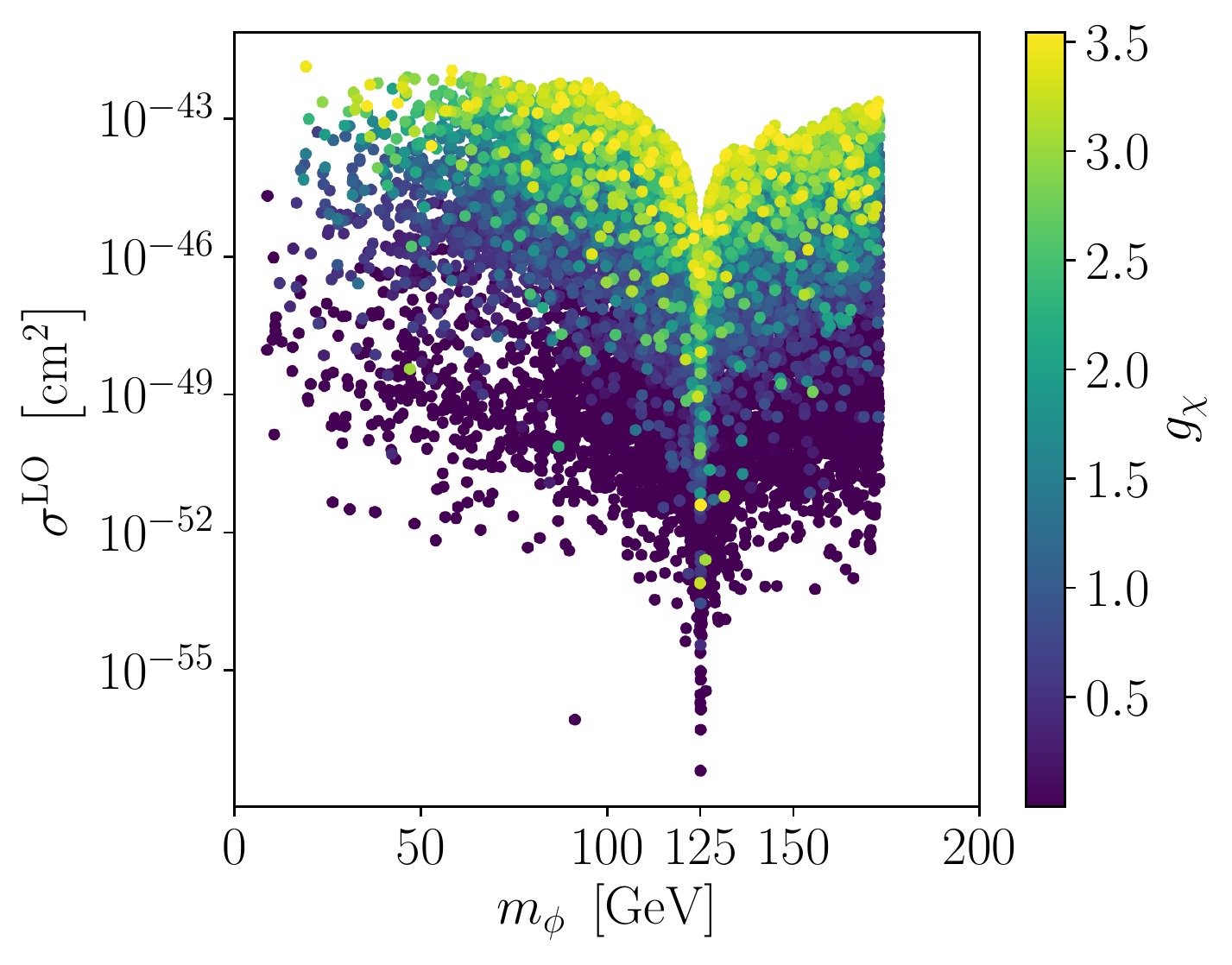}}
    \subfigure{\includegraphics[width=0.45\textwidth]{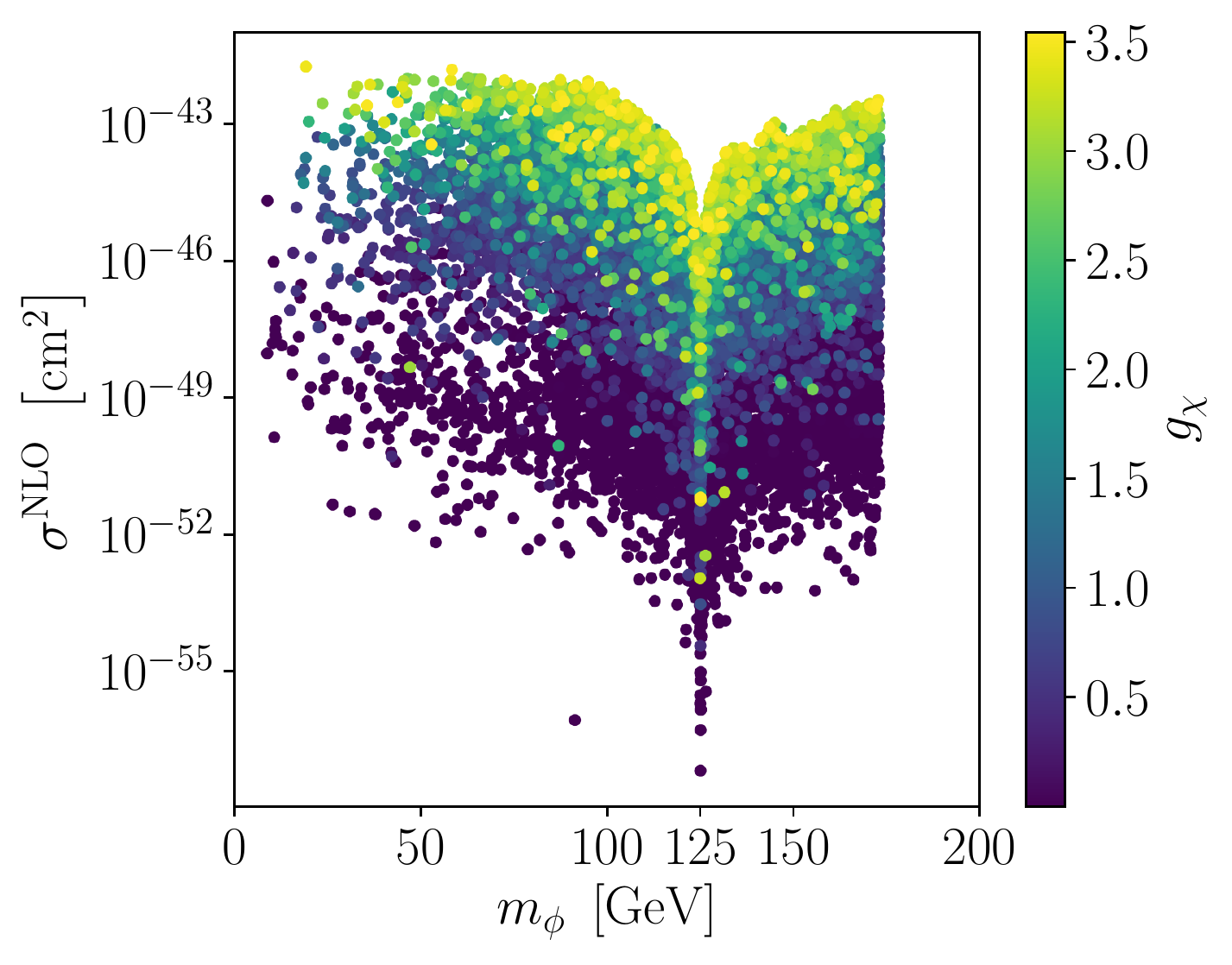}}
    \caption{Spin-independent direct detection LO cross section (left)
      and NLO cross section (right) versus the mass $m_\phi$ for the
      parameter sample passing all constraints and $m_\phi < m_t$. 
      The color code denotes the size of the dark gauge coupling
      $\gX$.}
    \label{fig:result2a}
\end{figure}

We now investigate the dependence of the LO and NLO direct detection
cross section on $\gX$ and the size of the NLO corrections for the
parameter sets featuring a non-SM-like Higgs boson with a mass $m_\phi
< m_t$. For these, the approximate treatment of the NLO box
contributions discussed in Subsection~\ref{sec:boxcorrections} can be
applied. In Fig.~\ref{fig:result2a} we display for all parameter sets
passing our constraints that additionally feature $m_\phi < m_t$ the
LO direct detection cross section in the left panel and the NLO result
in the right panel, as a function of $m_\phi$. The color code
quantifies the coupling $\gX \le \sqrt{4 \pi}$. Note,
  that here and in the following we do not
  apply the correction factor $f_{\chi\chi}$ Eq.~(\ref{eq:dmfraction})
  on the direct detection limit, as long as 
  we do not directly compare with the Xenon limit. This is why the LO
  cross section in Fig.~\ref{fig:result2a} is larger than in Fig.~\ref{result1}.\s

Both the LO and the NLO contribution to the SI direct detection cross
section are proportional to $f_q^{\text{LO}}$ and therefore
proportional to $\gX$, $\sin 2\alpha$ and  $(\mh^2 - \mH^2) \equiv
(m_h^2 -m_\phi^2)$. This behaviour is reflected in
Fig.~\ref{fig:result2a}. We observe that the LO cross section
increases with $\gX$, more specifically $\gX^2$ (yellow
points) and drops for $m_\phi = m_h = 125.09$~GeV.
The NLO corrections on the other hand increase with $g_\chi^3$. The
reason is that, as we explicitly verified, the NLO corrections are dominated by
the vertex corrections. The vertex corrections are proportional to $\gX^2$, so that the
NLO contribution to the cross section scales as $2\,\mbox{Re} (f_q^{\text{LO}}
f_q^{\text{vertex}*}) \propto \gX^3$, in contrast
to the LO contribution that is proportional to $\gX^2$. In total the
$K$-factor, {\it i.e.}~the ratio between NLO and LO cross section,
therefore increases with $\gX$. \s  

Being proportional to $f_q^{\text{LO}}$ the NLO corrected 
cross section also drops for $m_\phi = m_h$, so that the
sensitivity of the direct detection experiment is not increased after
inclusion of the NLO corrections; the blind spots remain also at
NLO. In our scan we furthermore did not find any parameter points
where a specific parameter combination leads to an accidental
suppression at LO that is removed at NLO. 
There is a further blind spot when $\alpha  = 0$. However, in this case the
SM-like Higgs boson has exactly SM-like couplings and the new scalar
decouples from all SM particles except for the coupling with the
SM-like Higgs boson. In this scenario the SM-like Higgs decouples from
the vector dark matter particle, and, depending on the mass of the
second scalar and its coupling strength with the SM-like Higgs boson,
we may end up with two dark matter candidates with the second scalar
being metastable. The study of such a scenario is beyond the scope of
this paper and we do not consider this case further here. \s

\begin{figure}[ht!]
    \centering
    \subfigure{\includegraphics[width=0.45\textwidth]{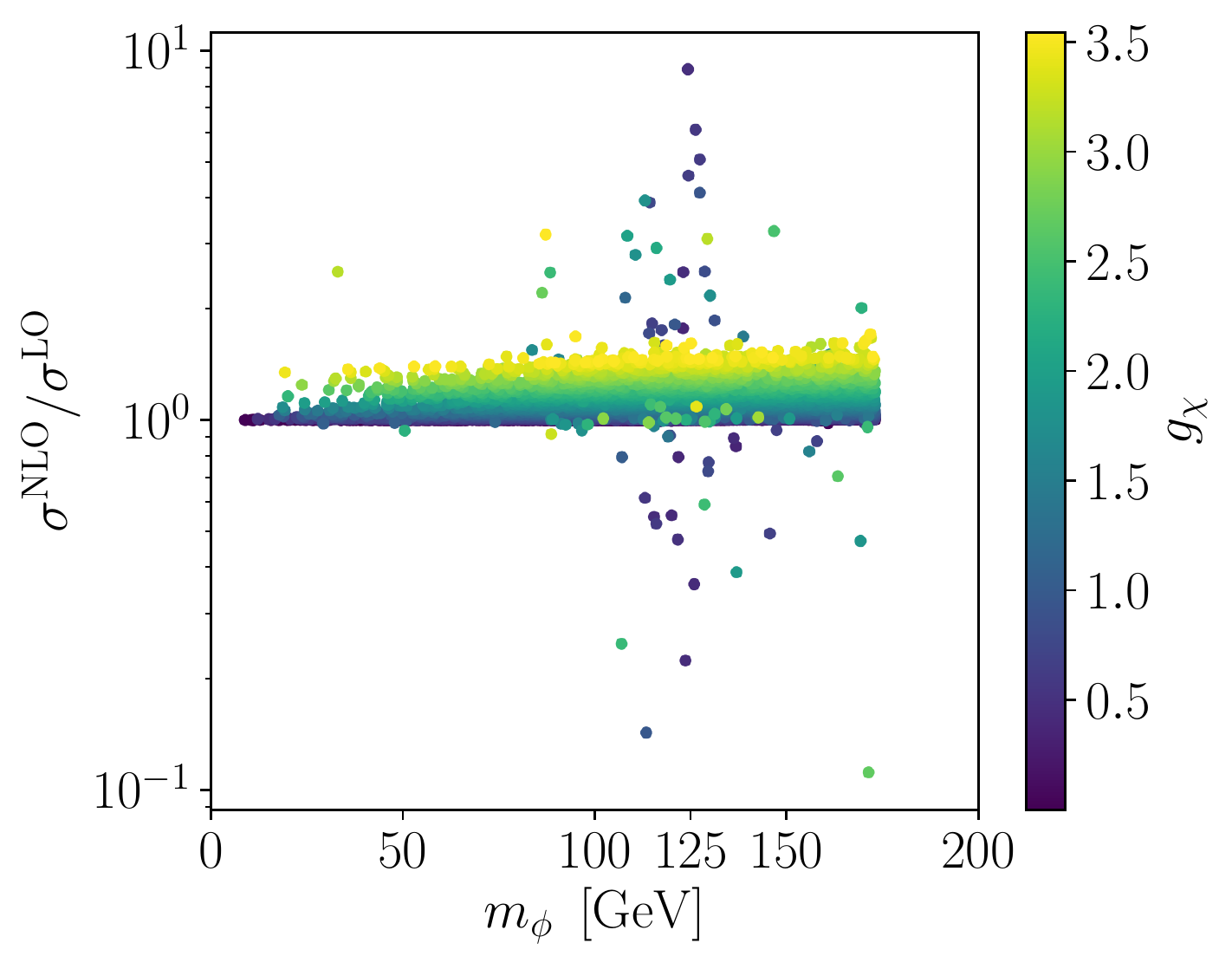}}   
    \subfigure{\includegraphics[width=0.45\textwidth]{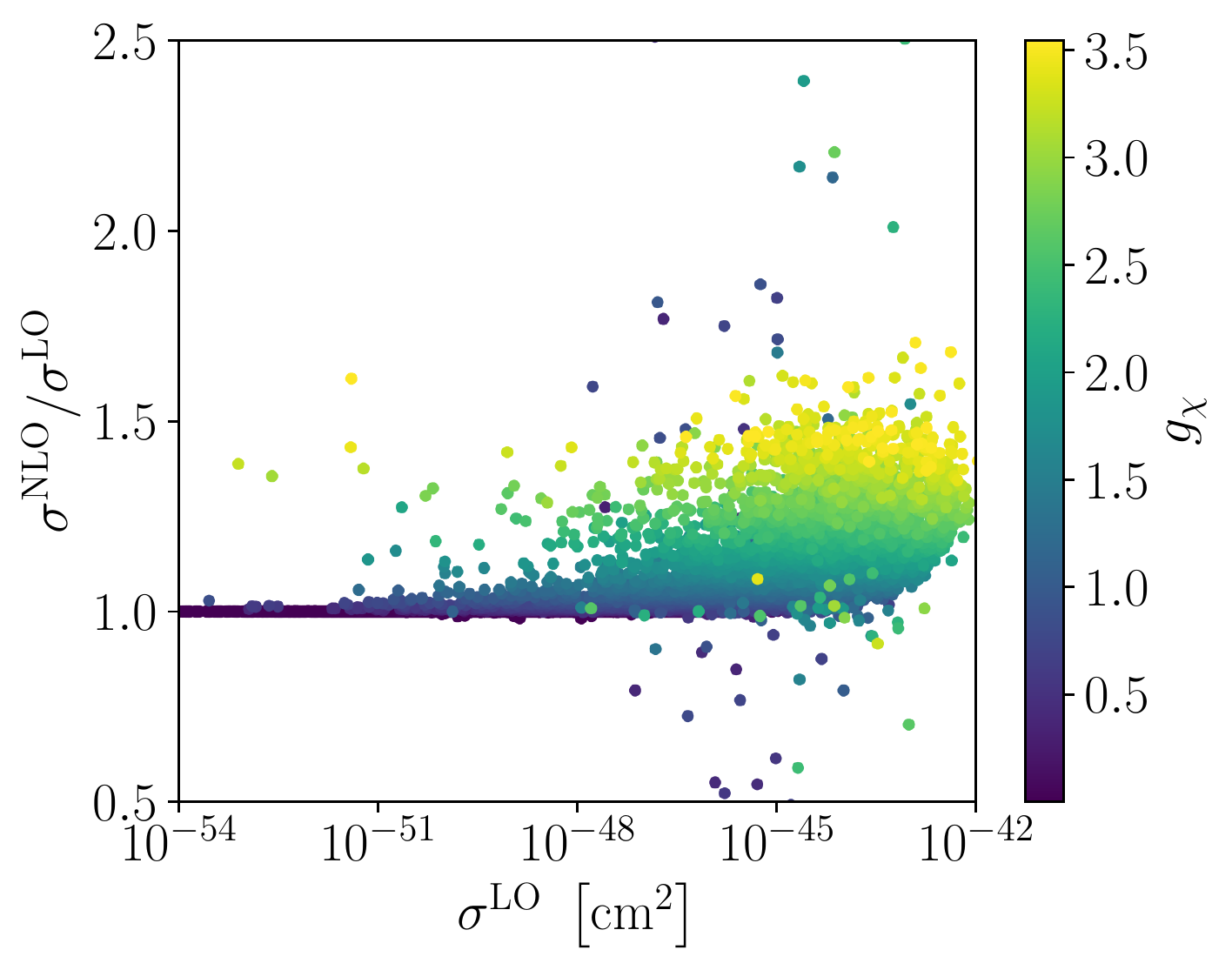}} 
    \caption{$K$-factor versus the Higgs mass $m_\phi$ (left) and
      $\sigma^{\text{LO}}$ (right) for the
      parameter sample passing all constraints and $m_\phi <
      m_t$. The color code denotes the size of the dark gauge coupling
      $\gX$.}
    \label{fig:result2b}
\end{figure}
The $K$-factor is depicted in Fig.~\ref{fig:result2b}, as a function of $m_\phi$
(left) and $\sigma^{\text{LO}}$ (right). The colour code indicates the
size of $\gX$. The $K$-factor is mostly positive and the bulk of $K$-facture values
ranges between 1 and about 2.3. As mentioned above, the $K$-factor
increases with $\gX$, as can also be inferred from the figure, in
particular from Fig.~\ref{fig:result2b} (right).  \s

In this and all other plots, we excluded points with $m_\phi \approx
m_h$ and $K$-factors where $|K| > 2.5$. We found that for $m_\phi
\approx m_h$ the interference effects between the $h$ and $\phi$
contributions, that become relevant here, largely increase the
(dominant) vertex contribution $f_q^{\text{vertex}}$ to the effective
NLO form factor. It exceeds by far the LO form factor
$f_q^{\text{LO}}$. Depending on the sign of $f_q^{\text{vertex}}$, the
NLO cross section, which is proportional to $2 \, \mbox{Re} (f_q^{\text{LO}}
f_q^{\text{vertex}*})$, is largely increased or suppressed, inducing
for large negative NLO amplitudes negative NLO cross sections. In these regions, the
NLO results are therefore no longer reliable. Two-loop contributions might
lead to a better perturbative convergence, but are beyond the scope of
this paper. We deliberately did not take into account one-loop squared
terms to remove the negative cross sections. Such an approach would only
include parts of the two-loop contributions. Whether or not they approximate the
total two-loop results well enough can only be judged after
performing the complete two-loop calculation. This is why we chose the
conservative approach to exclude these points from our analysis. \s

\begin{figure}[ht!]
    \centering
    \subfigure{\includegraphics[width=0.45\textwidth]{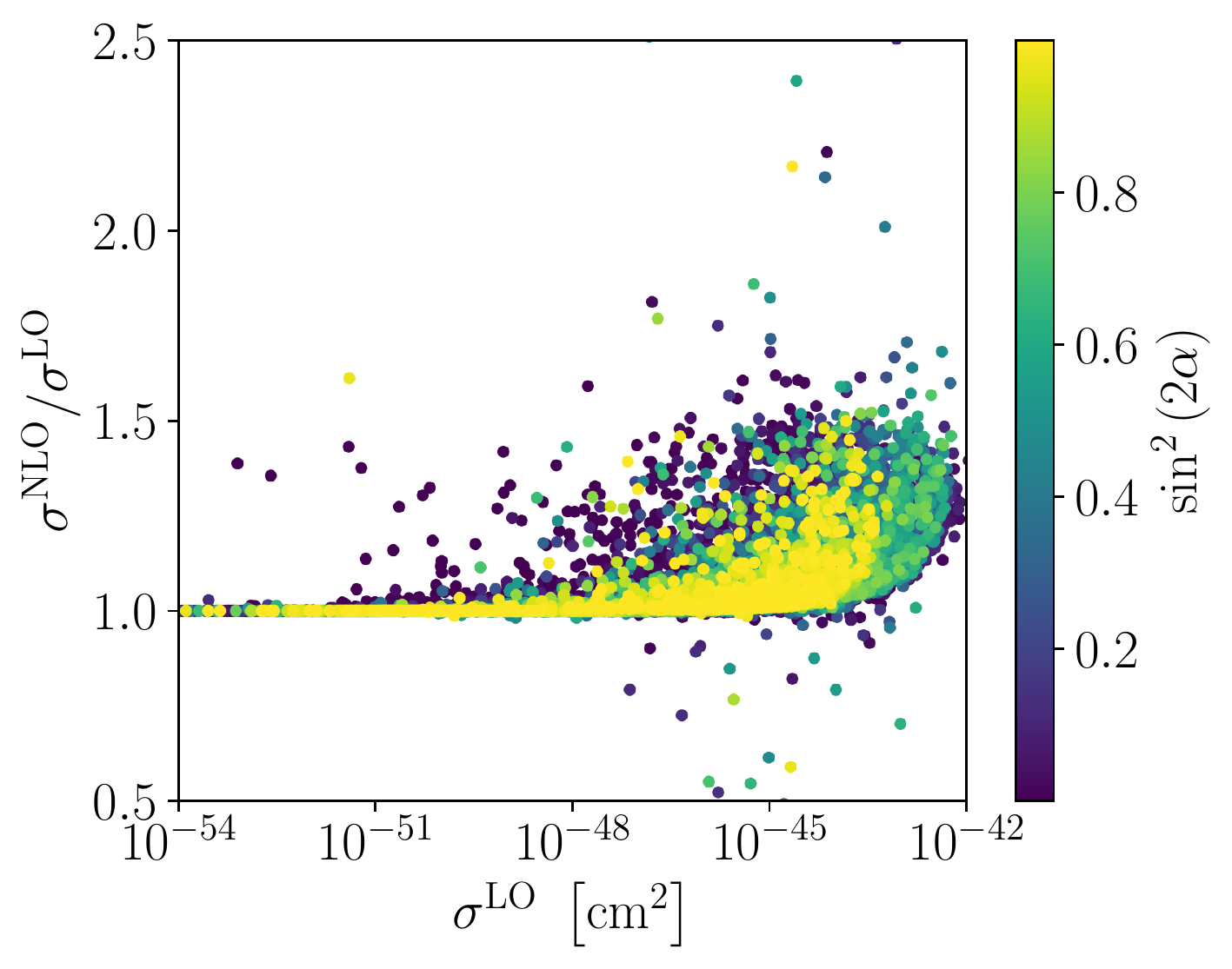}}   
    \subfigure{\includegraphics[width=0.45\textwidth]{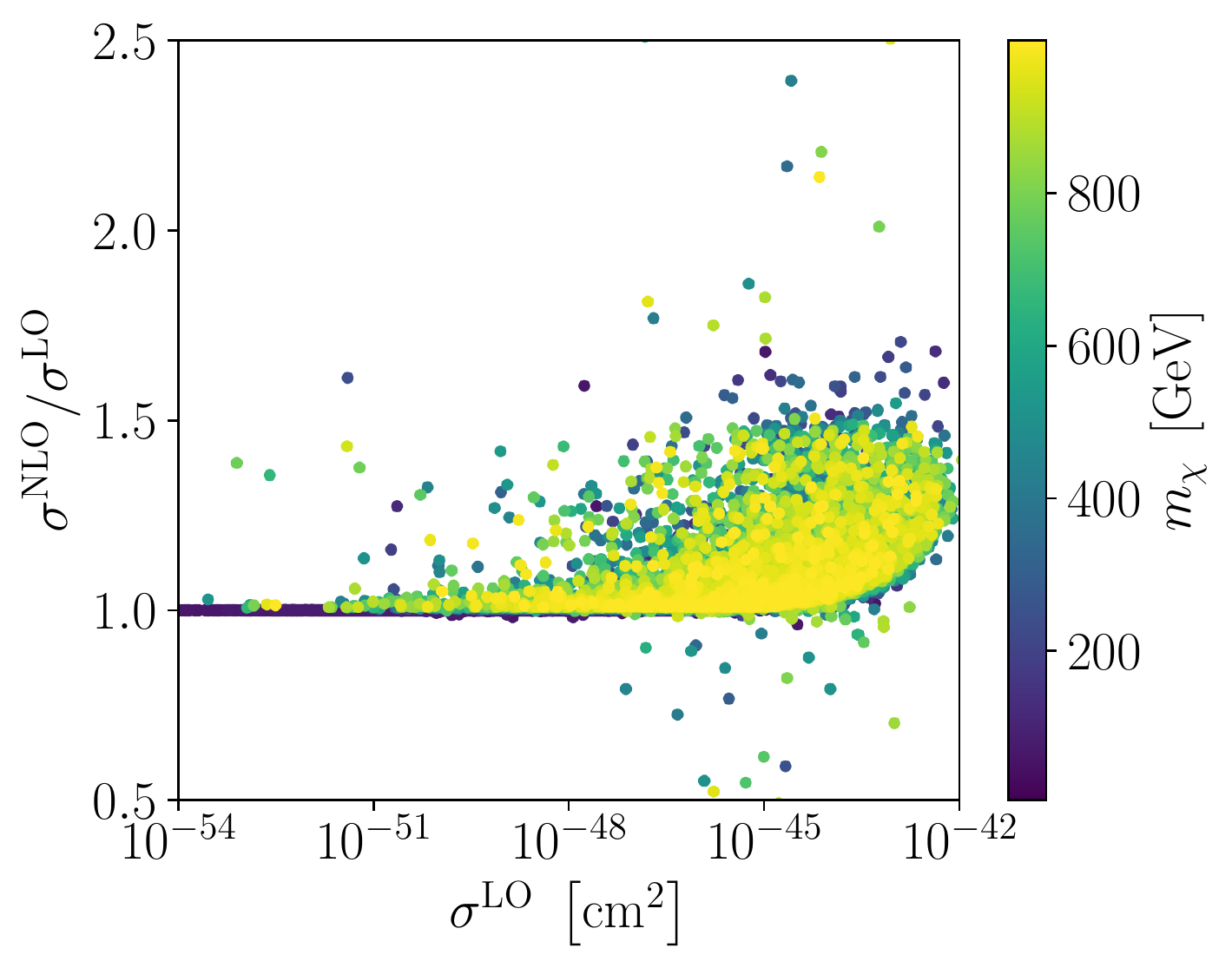}}  
    \caption{$K$-factor as function of the LO direct detection cross
      section with the color code indicating the size of
      $\sin^2 2\alpha$ (left) and $\mX$ (right).}
\label{fig:furtherkfactor}
\end{figure}

In Fig.~\ref{fig:furtherkfactor}, we show the $K$-factor as function of
$\sigma^{\text{LO}}$, but with the colour code indicating the size of
$\sin^2 2\alpha$ (left) and $\mX$ right. There is no
clear correlation between the $K$-factor and $\sin^2 2\alpha$ or
$m_\chi$. These plots furthermore show, that there is no correlation
between the maximum size of $\sigma^{\text{LO}}$ and $\mX$ or
$\sin^2 2\alpha$, while the maximum $\sigma^{\text{LO}}$ values require
large $\gX$ values, {\it cf.}~Fig.~\ref{fig:result2b} (right).

%%%%%%%%%%%%%%%%%%%%%%%%%%%%%%%%%%%%%%%%%%%%%%%%%%%%%%%%%%%%
\subsubsection{Results for $m_\phi > m_t$}
We now turn to the parameter region of our sample of valid points
where the approximation described in
Subsection~\ref{sec:boxcorrections} is a priori not valid. We cannot
judge the goodness of the approximation in this parameter region
without doing the full two-loop calculation which is beyond the scope
of this paper. We can check, however, if we see some unusual behaviour
compared to the results for parameter sets with $m_\phi < m_t$, where
the approximation can be applied. \s

\begin{figure}[t!]
    \centering
    \subfigure{\includegraphics[width=0.5\textwidth]{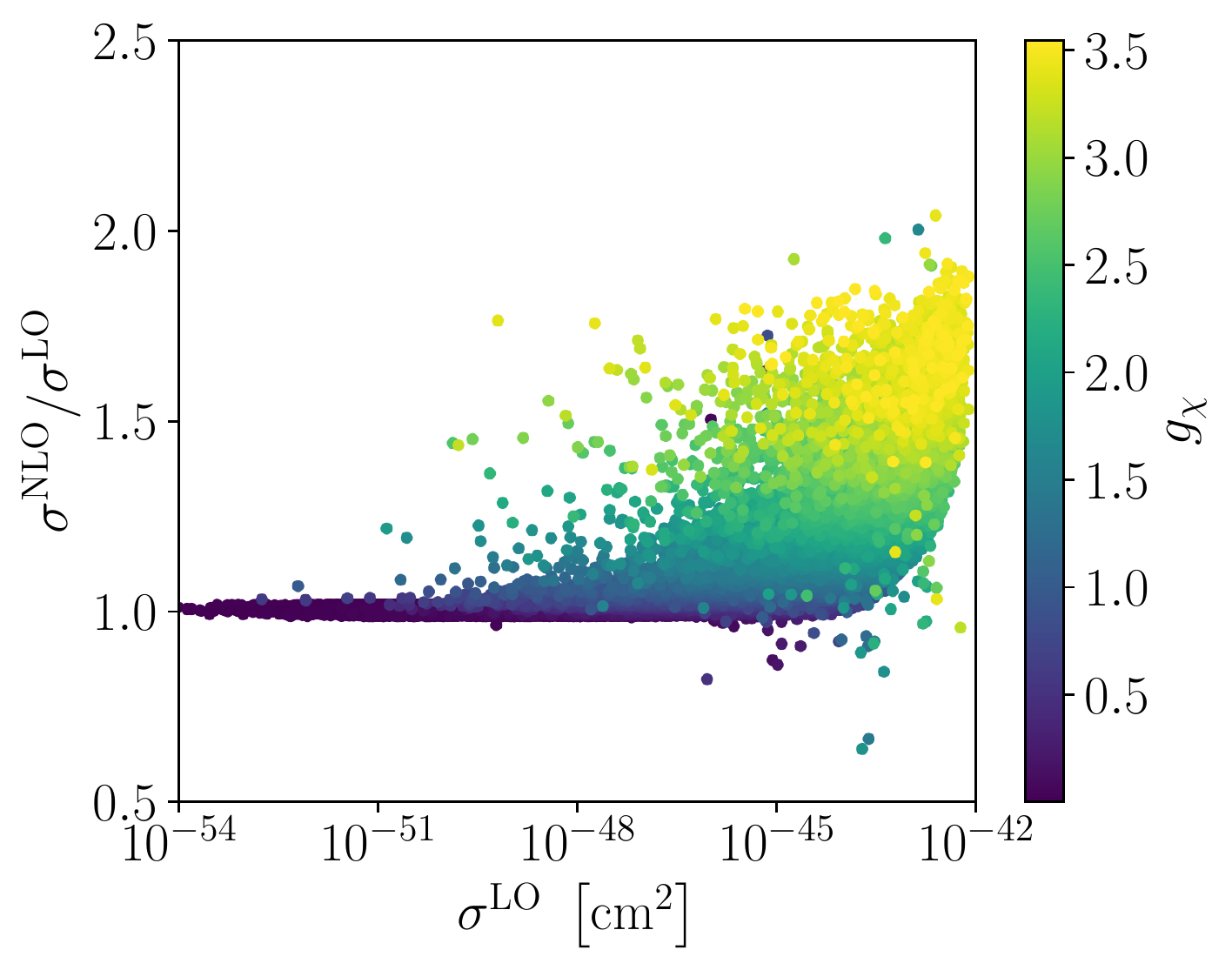}}
        \caption{$K$-factor versus the LO SI cross section. The color code 
      denotes the size of the dark gauge coupling $\gX$ for the
      parameter sample passing all constraints and $m_\phi >
      m_t$.}
    \label{fig:result3}
\end{figure}

Figure~\ref{fig:result3} shows the $K$-factor as a function of the LO
SI direct detection cross section. The size of $\gX$ is indicated by the color
code. We only take into account parameter samples compatible with all
constraints and where $m_\phi >m_t$. As already observed and discussed
for the parameter sample with $m_\phi < m_t$, also here the $K$-factor
increases with $\gX$. Overall, the bulk of points reaches larger
$K$-factors than for $m_\phi < m_t$ but remains below 2.5. So, the
behaviour of the $K$-factor does not substantially differ from the
results for $m_\phi < m_t$. The comparison of the approximate and exact
result in Ref.~\cite{Ertas:2019dew} showed that the difference in the box
contribution between the two results does not exceed one order of
magnitude for a pseudoscalar mediator with mass 1~TeV\footnote{We
  estimate this by extrapolating Fig.~4 (left) in
  Ref.~\cite{Ertas:2019dew} to 1~TeV} and remains small even for
scalar mediator masses up to 1~TeV ({\it cf.}~Fig.~4 in
Ref.~\cite{Ertas:2019dew}). Together with the fact that the box
contribution makes up only for a small part of the NLO SI direct
detection cross section\footnote{We explicitly verified that the box
  form factor $f_q^{\text{box}}$ remains below the vertex correction
  form factor $f_q^{\text{vertex}}$. In particular, for $K$-factors
  above 1, the box form factor remains more than two orders of
  magnitude below the vertex form factor.}, we conclude that our
approximate NLO results for parameter sets with larger mediator masses should also
be applicable in this parameter region. 

%%%%%%%%%%%%%%%%%%%%%%%%%%%%%%%%%%%%%%%%%%%%%%%%%%%%%%%%%%%%
\subsubsection{Gauge Dependence}
As has been discussed in Ref.~\cite{Krause:2016oke} for the 2HDM and
in Ref.~\cite{Krause:2017mal} for the N2HDM the renormalisation of the
mixing angle $\alpha$ 
in the KOSY scheme leads to gauge parameter dependent results. We
therefore check here the gauge dependence of our NLO results by
performing the calculation in the general $R_\xi$ gauge and comparing
it with our default result in the Feynman gauge $\xi=1$. \s

\begin{figure}[ht!]
    \centering
    \subfigure{\includegraphics[width=0.4\textwidth]{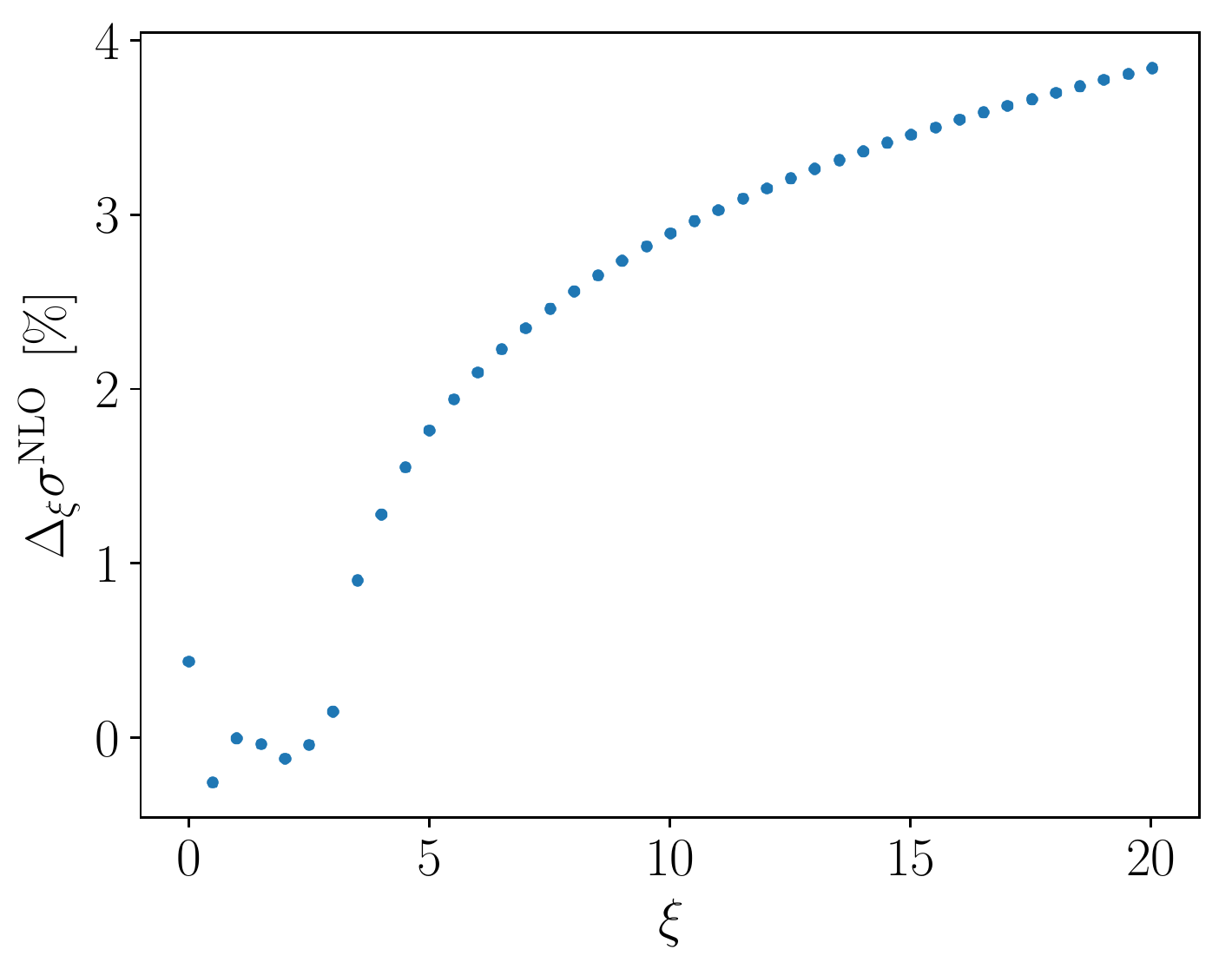}}
    \subfigure{\includegraphics[width=0.4\textwidth]{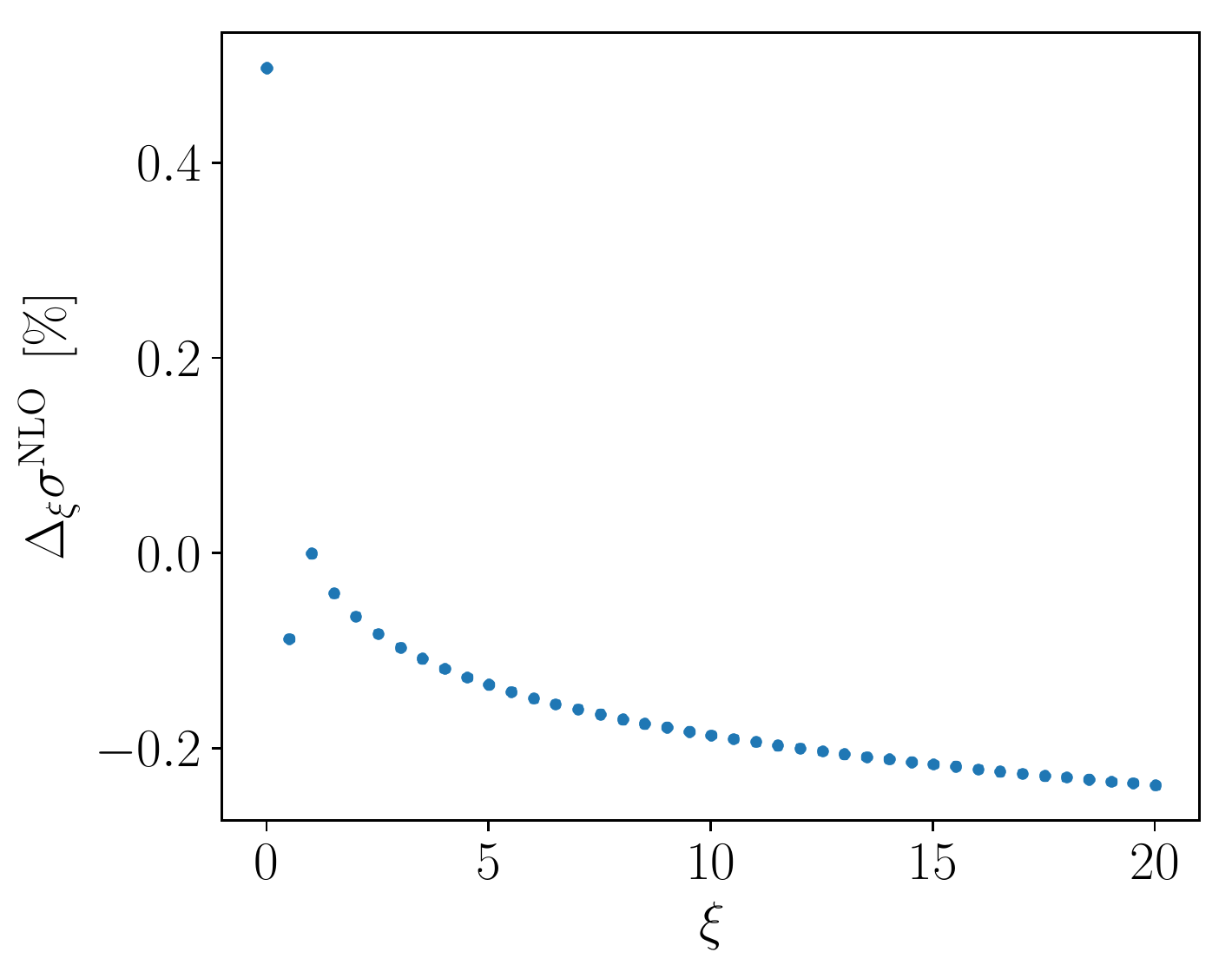}}
    \caption{Relative gauge dependence $\Delta_{\xi}\sigma$ versus the
      gauge parameter $\xi$ for parameter point number 5 (left) and 6
      (right). See text, for their definitions. }
    \label{fig:result4}
\end{figure}

We introduce the relative gauge dependence $\Delta_\xi \sigma^{\text{NLO}}$, defined as 
\beq 
\Delta_{\xi}\sigma^{\text{NLO}} =
\cbrak{\sigma^{\text{NLO}}\big\vert_{\xi}-\sigma^{\text{NLO}}\big\vert_{\xi=1}}/\sigma^{\text{NLO}}\big\vert_{\xi=1}  \;,
\eeq
where $\sigma^{\text{NLO}}\big\vert_{\xi}$ denotes the NLO SI direct
detection cross section calculated in the general $R_\xi$ gauge and 
$\sigma^{\text{NLO}}\big\vert_{\xi=1}$ the result in the Feynman
gauge. In Fig.~\ref{fig:result4} we show $\Delta_\xi
\sigma^{\text{NLO}}$ as a function of the gauge parameter $\xi$, for
two sample parameter points of our valid parameter set, called point 5
and 6, respectively. They are given by the following input
parameters. For the parameter point 5 we have
\beq
\begin{array}{lllllll}
\mbox{\underline{Point 5: }} \; & m_\phi &=& 283.44 \mbox{ GeV }\;, & 
\mX &=& 914.76 \mbox{ GeV } \;, \\
& \gX &=& 7.67 \;, & \alpha &=& 0.07312 \;.
\end{array}
\eeq
The parameter point 6 is given by 
\beq
\begin{array}{lllllll}
\mbox{\underline{Point 6: }} \; & m_\phi &=& 119.84 \mbox{ GeV }\;, & 
\mX &=& 766.82\mbox{ GeV } \;, \\
& \gX &=& 1.555  \;, & \alpha &=& 0.425943 \;.
\end{array}
\eeq

As can be inferred from Fig.~\ref{fig:result4}, we clearly see a gauge
dependence of the NLO results. The relative gauge dependence is,
however, small with values below the few per cent level for a rather large
range of $\xi$ variation. Note also, that a gauge parameter dependence
a priori is no problem as long as it is made sure that the explicit
value of the gauge parameter is accounted for when interpreting the
results. 

%%%%%%%%%%%%%%%%%%%%%%%%%%%%%%%%%%%%%%%%%%%%%%%%%%%%%%%%%%%%%
\subsubsection{Renormalisation Scheme Dependence} 
\begin{figure}[ht!]
    \centering
    \includegraphics[width=0.5\textwidth]{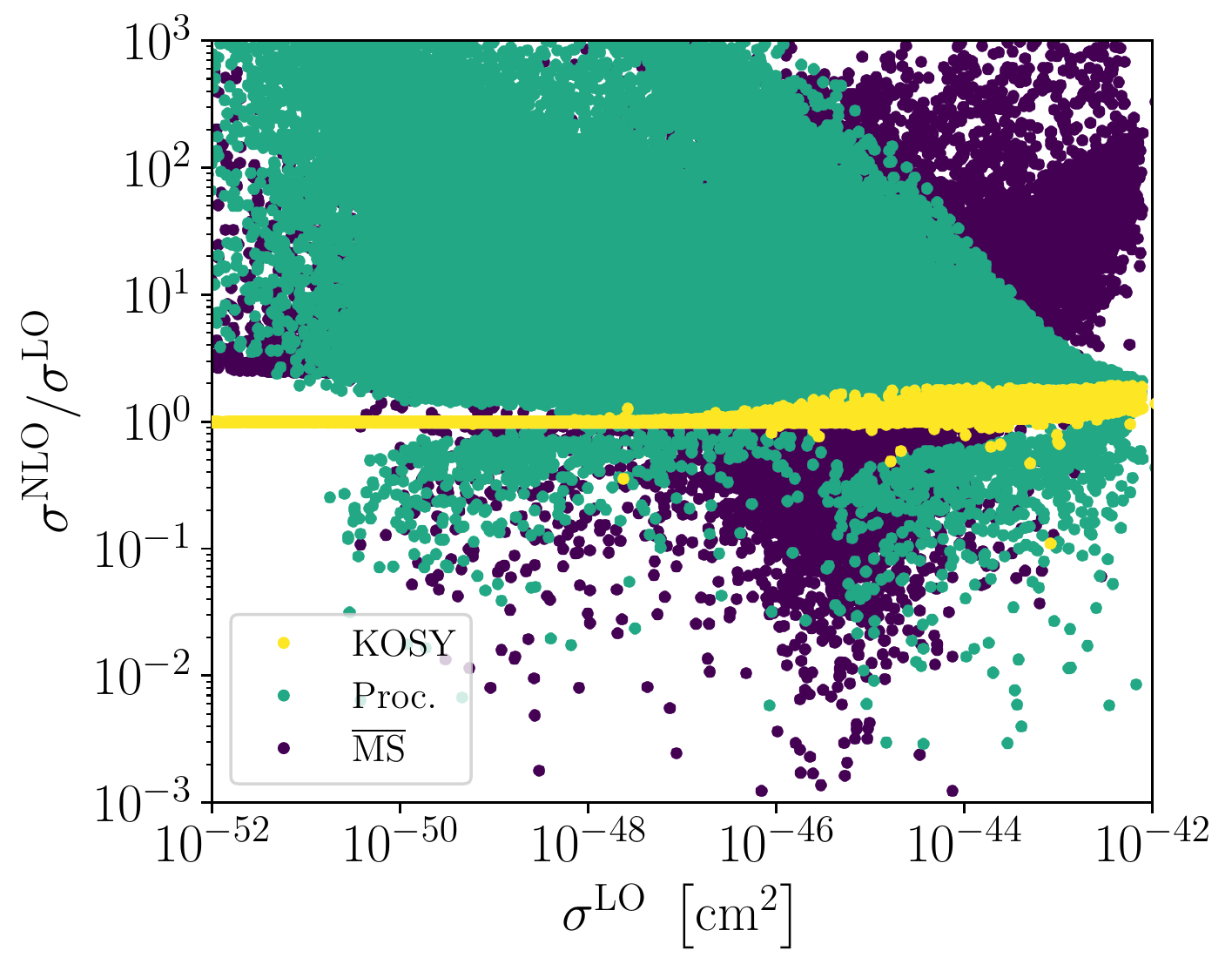}
    \caption{The $K$-factor versus the LO SI direct detection cross
      section for the whole data sample passing all constraints and
      for three different renormalisation schemes of $\alpha$: the KOSY 
      scheme (yellow), the process-dependent scheme (green), the
      $\overline{\mbox{MS}}$ scheme (violet).}
    \label{fig:result5}
\end{figure}

We now investigate the influence of the renormalisation scheme and
scale on the NLO result. For this, we show in Fig.~\ref{fig:result5}
the $K$-factor for the whole data sample passing our constraints for
three different renormalisation schemes of the mixing angle $\alpha$
as a function of the LO cross section. The chosen schemes have been
described in Subsection~\ref{sec:alpharenorm} and are the KOSY scheme
(yellow points), the process-dependent scheme (green) and the
$\overline{\mbox{MS}}$ scheme (violet). The scale applied in the
$\overline{\mbox{MS}}$ scheme is $\mu_0 = 1$~GeV.
The KOSY scheme has been shown
to lead to a gauge-parameter dependent definition of the counterterm
$\delta\alpha$~\cite{Krause:2016oke,Krause:2017mal}. This is also the
case for the $\overline{\mbox{MS}}$ scheme. As can be inferred from the plot,
the $\overline{\mbox{MS}}$ scheme additionally leads to unnaturally large NLO
corrections with $K$-factor values up to about $10^8$ for our data
sample (not shown in the plot). This has been known already
from previous investigations in the 2HDM \cite{Krause:2016oke} and
N2HDM \cite{Krause:2017mal}. The process-dependent scheme has the
virtue of implying a manifestly gauge-independent definition of the
mixing angle counterterm. However, also here the NLO corrections are
unacceptably large with values up to about $10^9$, so that also this
scheme turns out to be unsuitable for practical use. This behaviour
has also been observed in our previous works
\cite{Krause:2016oke,Krause:2017mal}. We therefore conclude that the
KOSY scheme should be used in the computation of the NLO
corrections. The fact that it is gauge dependent is no problem as long
as the chosen gauge is clearly stated when presenting
results. Moreover, by applying a pinched scheme, the gauge dependence
can be avoided, {\it cf.}~Ref.~\cite{Krause:2016oke}. This is left for
future work. 
\s 

The uncertainty due to missing higher-order corrections can be 
estimated by varying the renormalisation scheme or by varying the 
renormalisation scale. The comparison of the KOSY with the other two
renormalisation schemes makes no sense as the latter lead to
unacceptably large corrections. The KOSY scheme does not allow us to vary the
renormalisation scale, so that we cannot provide an estimate of the
uncertainty due to missing higher order corrections. We conclude with
the remark that the variation of the renormalisation scale between 1/2
and 2 times the scale $\mu_0$ in the $\overline{\mbox{MS}}$ scheme
leads to a variation of the NLO cross section of about
16\% - in contrast to the unphysically large corrections that are to
be traced back to the blowing-up of the $\overline{\mbox{MS}}$
counterterm of $\alpha$. 

%%%%%%%%%%%%%%%%%%%%%%%%%%%%%%%%%%%%%%%%%%%%%%%%%%%%%%%%%%%%%%
\subsubsection{Phenomenological Impact of the NLO Corrections on the Xenon Limit}
\begin{figure}[ht!]
    \centering
    \subfigure{\includegraphics[width=0.45\textwidth]{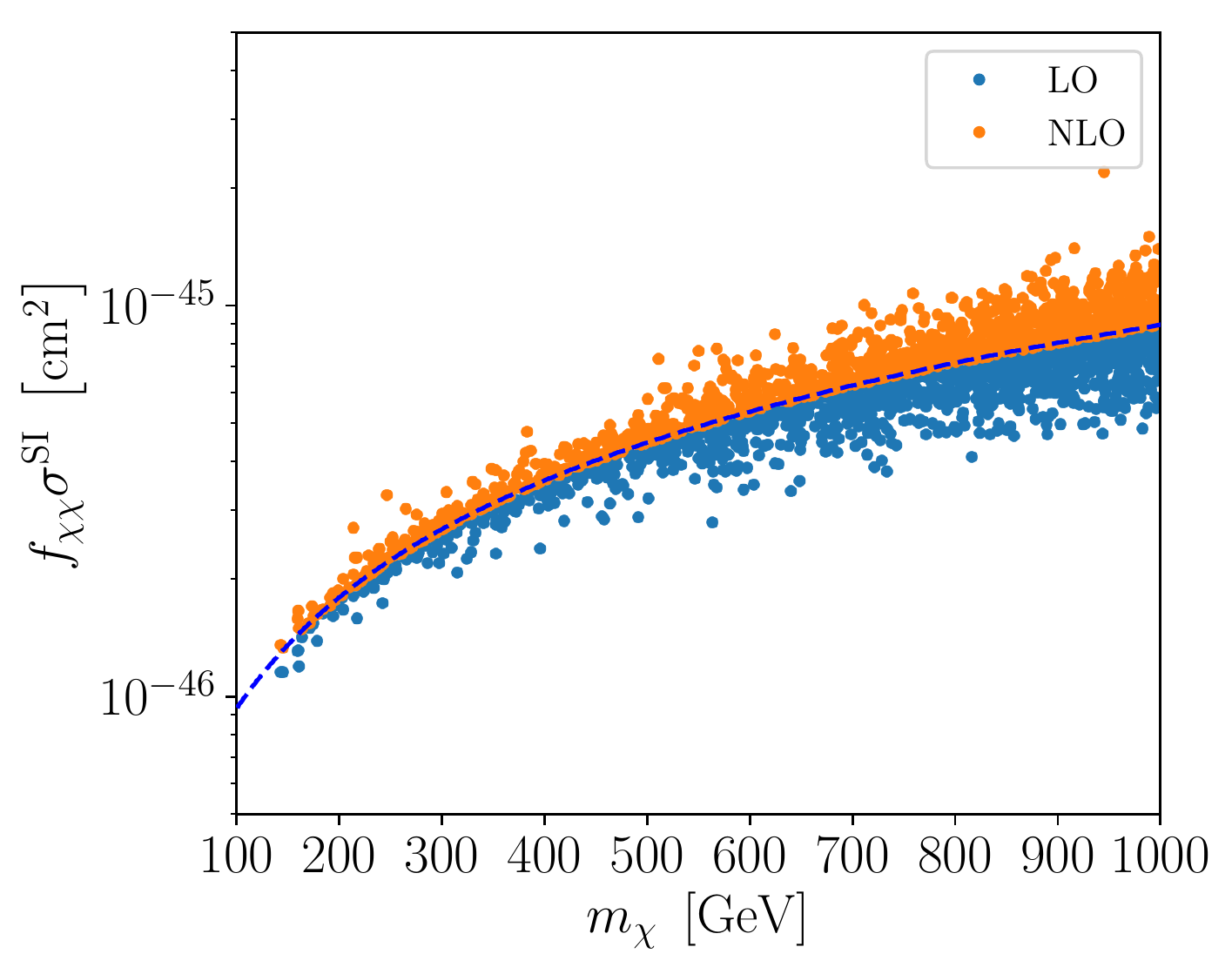}}
    \subfigure{\includegraphics[width=0.45\textwidth]{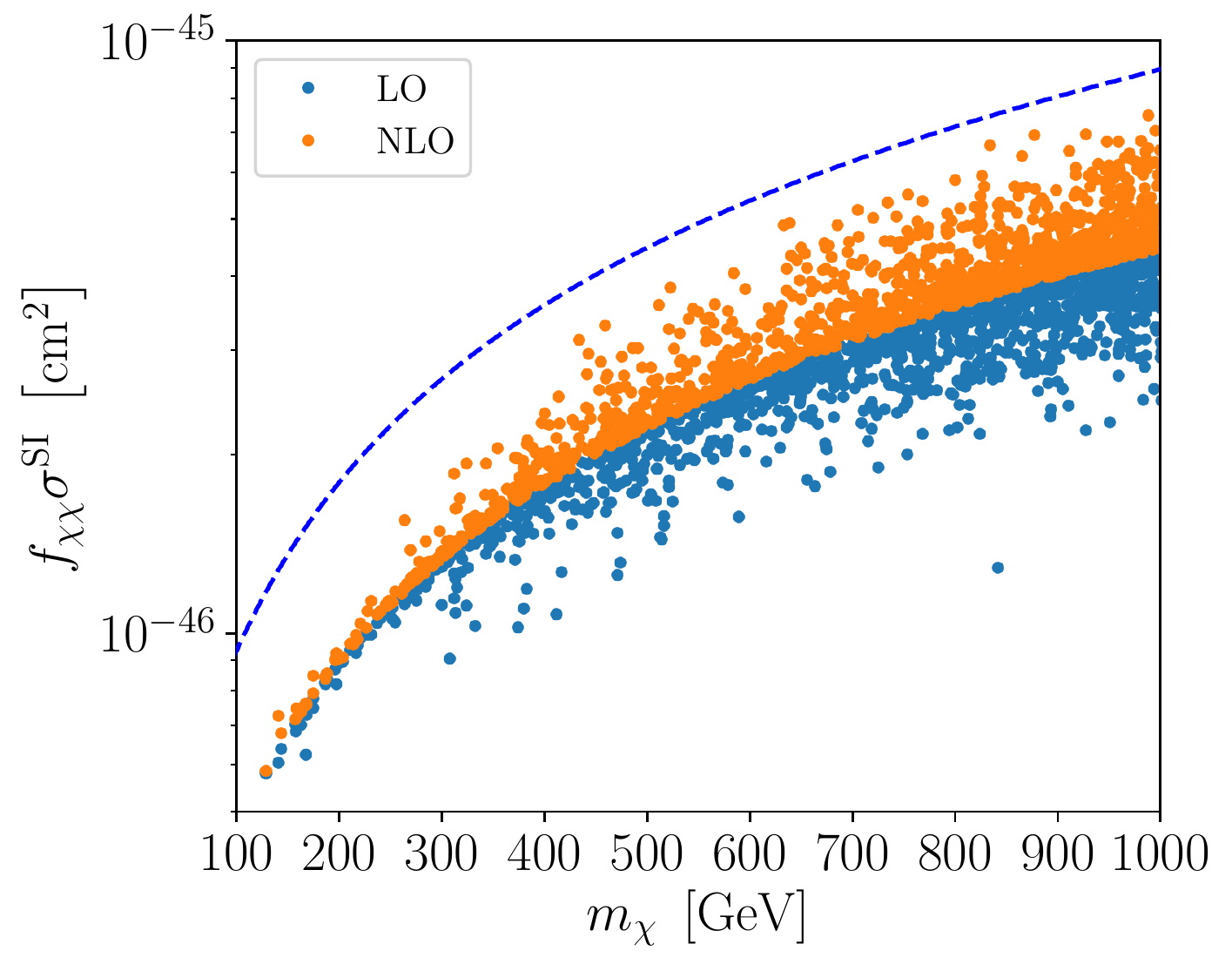}}
    
    \caption{The SI cross section including the correction factor
      $f_{\chi\chi}$ at LO (biue) and NLO (orange) compared to the
      Xenon limit (blue-dashed) versus the DM mass $\mX$. The definition
    of the parameter sample included in the left and right plots is described in the text.}
    \label{fig:result6}
\end{figure}

We now turn to the discussion of the phenomenological impact of our
NLO results. In Fig.~\ref{fig:result6} (left) we show the LO direct
detection cross section (blue points) and the NLO result (orange)
compared to the Xenon limit (blue-dashed), as a function of the DM particle
mass. For the consistent comparison with the Xenon limit we applied
the correction factor $f_{\chi\chi}$ (Eq.~\ref{cor_factor}) to the
LO and NLO cross section in both plots of Fig.~\ref{fig:result6}. In
the left figure we plot all parameter points where the LO cross
section does not exceed the Xenon limit but the NLO result does. This
plot shows that for a sizeable number of parameter points, the
compatibility with the experimental constraints would not hold at NLO
any more. This demonstrates that the NLO corrections are 
important and need to be accounted for in order to make reliable
predictions about the viable parameter space of the VDM model. \s

In the right plot we display the same quantities, but only for
parameter points of our data sample where 
\begin{equation}
        \frac{\vert \sigma_{{\text{Xe}}}(\mX)-\sigma^{\text{LO}}\vert}{\sigma^{\text{LO}}}>1
\end{equation}
and 
\begin{equation}
        \frac{\vert
          \sigma_{\text{Xe}}(\mX)-\sigma^{\text{NLO}}\vert}{\sigma^{\text{NLO}}}<
        1 \;.
\end{equation}
This implies we only consider parameter points where the LO cross
section is much smaller than the Xenon limit, but the NLO cross
section is of the order of the Xenon limit. We learn from this figure
that although LO results might suggest that the Xenon experiment is
not sensitive to the model, this statement does not hold any more when
NLO corrections are taken into account. These results confirm the
importance of the NLO corrections when interpreting the data. 

%%%%%%%%%%%%%%%%%%%%%%%%%%%%%%%%%%%%%%%%%%%%%%%%%%%%%%%
\section{Conclusions}\label{sec:conclusion}
In this paper, we investigated a minimal model with a VDM particle. 
We computed the NLO corrections to the direct detection cross section
for the scattering of the VDM particle off a nucleon. We developed
the renormalisation of the model, proposing several renormalisation
schemes for the mixing angle $\alpha$ of the two physical scalars of
the model. We computed the leading corrections,
including relevant two-loop box contributions to the effective gluon
interaction in the heavy quark approximation. With the box
contributions to the NLO cross section being two orders of magnitude
below the leading vertex corrections, we estimated the
error induced by the approximation to be small. Interference effects
of the two scalar particles that become important for degenerate mass
values on the other hand, were found to be large and require further
investigations beyond the 
scope of this paper, namely the computation of the complete two-loop
contributions. Outside this region, the perturbative series is
well-behaved and $K$-factors of up to about 2.5 were found. 
\s

We further investigated the impact of the chosen renormalisation scheme for
$\alpha$. While the process-dependent renormalisation of $\alpha$ is
manifestly gauge-parameter independent, it was found to lead to
unphysically large corrections. This did not improve by choosing the
gauge-parameter dependent 
$\overline{\mbox{MS}}$ scheme. A renormalisation scheme exploiting the
OS conditions of the scalar fields on the other hand, leads to moderate
$K$-factors, while being manifestly gauge-parameter dependent. For the
proper interpretation of the data, therefore, the choice of the gauge
parameter has to be specified here. \s 

We found that the NLO corrections can either enhance or suppress the
cross section. With $K$-factors of up to about 2.5, they are important
for the correct interpretation of the  viability of the VDM model 
based on the experimental limits on the  
direct detection cross section. The NLO corrections can increase the 
LO results to values where the Xenon experiment becomes sensitive to
the model, or to values where the model is even excluded due to cross sections
above the Xenon limit. In case of 
suppression, parameter points that might be rejected at LO may render 
the model viable when NLO corrections are included. \s

The next steps would be to investigate in greater detail the
interesting region of degenerate scalar masses and study its implication on
phenomenology in order to further be able to delineate the viability of this
simple SM extension in providing a VDM candidate. 

%%%%%%%%%%%%%%%%%%%%%%%%%%%%%%%%%%%%%%%%%%%%%%%%%%%%%%%
\subsubsection*{Acknowledgments}
We are thankful to M.~Gabelmann, M.~Krause and
M.~Spira for fruitful and clarifying discussions. We are grateful to
D.~Azevedo for providing us with the data samples. R.S. is supported
in part by a CERN grant CERN/FIS-PAR/0002/2017, an FCT grant
PTDC/FISPAR/31000/2017, by the CFTC-UL strategic project
UID/FIS/00618/2019 and by the NSC, Poland, HARMONIA
UMO-2015/18/M/ST2/00518. 
\vspace*{0.5cm}
%%%%%%%%%%%%%%%%%%%%%%%%%%%%%%%%%%%%%%%%%%%%%%%%%%%%%%%
\appendix
\section{Nuclear Form Factors}\label{APP::NUCLEAR}
We here present the numerical values for the nuclear form factors
defined in \cref{HISANO::MATRIXELEMENT}. The values of the form
factors for light quarks are taken from {\tt
  micrOmegas}\cite{Belanger:2018mqt} 
\begin{subequations}    
    \begin{align}
        & f^p_{T_u} = 0.01513\,,\quad f^p_{T_d} = 0.0.0191\,,\quad f^p_{T_s} = 0.0447\,, \\
        & f^n_{T_u} = 0.0110\,,\quad f^n_{T_d} = 0.0273\,,\quad f^n_{T_s} = 0.0447\,, 
    \end{align}
\end{subequations}
which can be related to the gluon form factors as
\begin{align}
    f^p_{T_G} = 1-\sum_{q=u,d,s} f_{T_q}^p\,,\qquad f^n_{T_G} = 1-\sum_{q=u,d,s} f_{T_q}^n\,.
\end{align}
The needed second momenta in \cref{HISANO::MATRIXELEMENT} are defined
at the scale $\mu=m_Z$ by using the {\tt CTEQ} parton distribution
functions \cite{Pumplin:2002vw},
\begin{subequations}    
    \begin{align}
        u^p(2) = 0.22\,,\qquad & \bar u^p(2) = 0.034\,, \\
        d^p(2) = 0.11\,,\qquad & \bar d^p(2) = 0.036\,, \\
        s^p(2) = 0.026\,,\qquad & \bar s^p(2) = 0.026\,, \\
        c^p(2) = 0.019\,,\qquad & \bar c^p(2) = 0.019\,, \\
        b^p(2) = 0.012\,,\qquad & \bar b^p(2) = 0.012\,, 
    \end{align}
\end{subequations}
where the respective second momenta for the neutron can be obtained by
interchanging up- and down-quark values. 
\section{Feynman Rules}
In the following we list the Feynman rules needed to perform the
one-loop calculation. The Feynman rules are derived by using the
program package {\tt
  SARAH}~\cite{Staub:2013tta,Staub:2012pb,Staub:2010jh,Staub:2009bi}. All
momentum conventions are adopted from the {\tt FeynArts} conventions. 
The trilinear Higgs couplings read
\begin{subequations}
  \begin{align}
      &\begin{aligned}
          \hbox{\includegraphics[width = 0.2\textwidth]{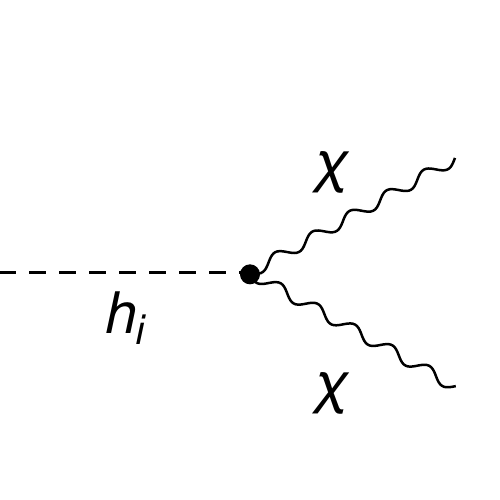}}
      \end{aligned}
       &=\ii 2 g_{\chi}m_{\chi}R_{\alpha,i2}\,,\quad
       &\begin{aligned}
          \includegraphics[width = 0.2\textwidth]{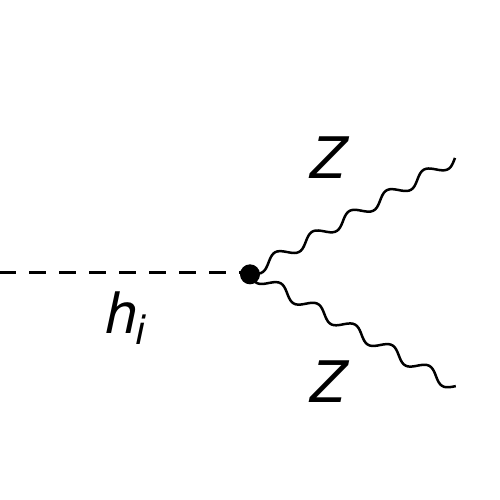} 
       \end{aligned}
       &=\ii \frac{g m_Z^2}{m_W} R_{\alpha , 1i}\,,\quad\\
       &\begin{aligned}
          \includegraphics[width = 0.2\textwidth]{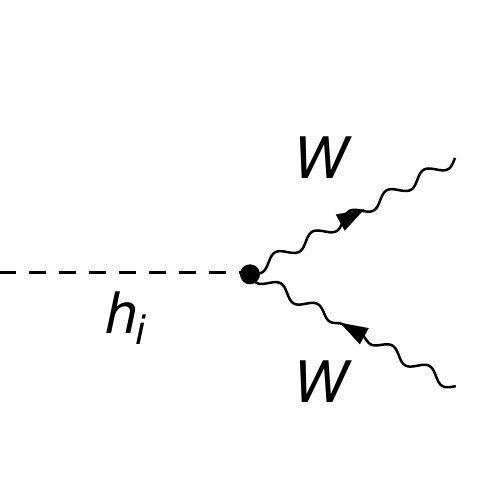} 
       \end{aligned}
       &=\ii g m_W R_{\alpha, 1i}\,,\quad
       &\begin{aligned}
          \includegraphics[width = 0.2\textwidth]{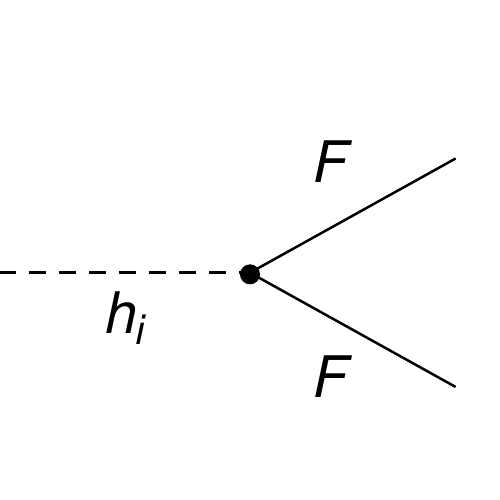}
       \end{aligned}
       &=\ii \frac{g m_F}{ 2 m_W} R_{\alpha,1i}\,,\quad\\
       &\begin{aligned}
          \includegraphics[width = 0.2\textwidth]{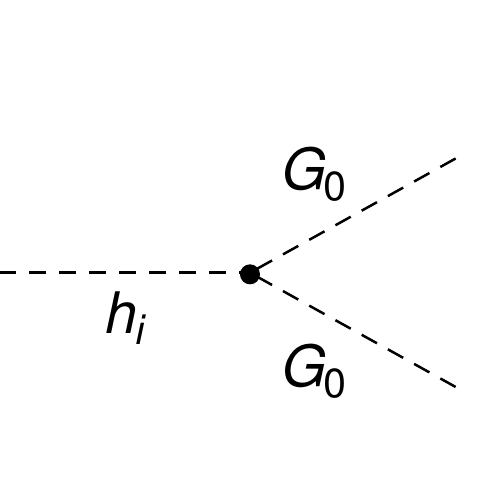}
       \end{aligned}
       &=-\ii\frac{g_{\chi} m_{h_i}^2}{m_{\chi}} R_{\alpha, i1}\,,\quad
       &\begin{aligned}
          \includegraphics[width = 0.2\textwidth]{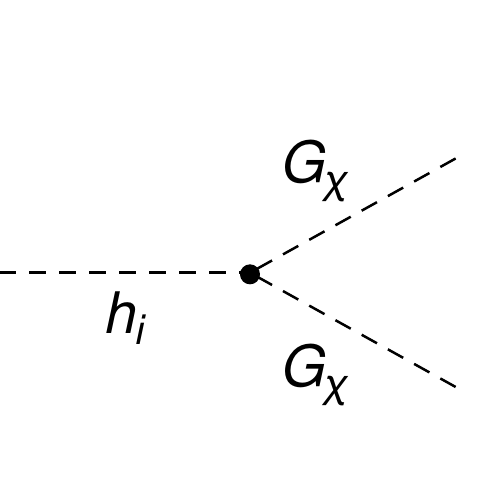}
       \end{aligned}
       &=-\ii\frac{g_{\chi} m_{h_i}^2}{m_{\chi}} R_{\alpha, i2}\,,\quad\\
       &\begin{aligned}
          \includegraphics[width = 0.2\textwidth]{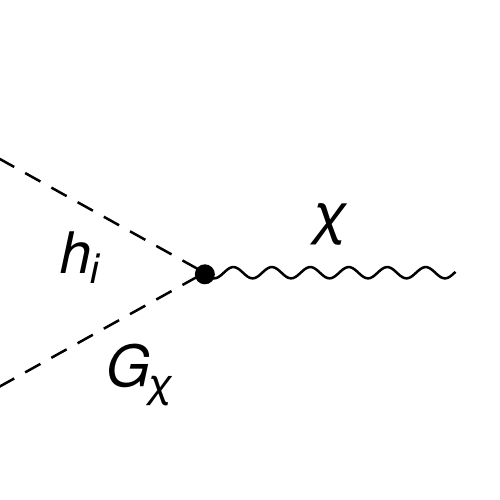}
       \end{aligned}
       &=- g_{\chi}(p_{G_{\chi}}-p_{h_i})R_{\alpha,1i}\,,\quad
       &\begin{aligned}
          \includegraphics[width = 0.2\textwidth]{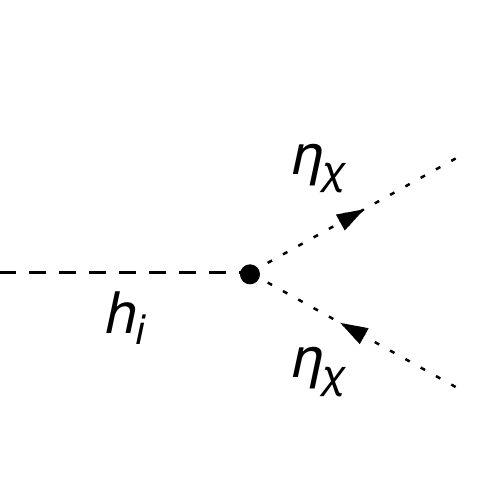}
       \end{aligned}
       &=\ii g_{\chi}m_{\chi}R_{\alpha,i2} \xi_{\chi}\,,\quad
  \end{align}
  \begin{align}
      \begin{aligned}
          \includegraphics[width = 0.2\textwidth]{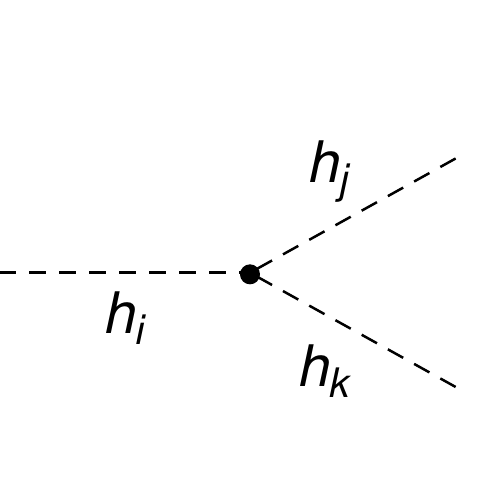}
      \end{aligned}
      =
      \begin{aligned}   
          \ii \left[\kappa v \left(R_{\alpha,1i}R_{\alpha,2j}R_{\alpha,2k}+R_{\alpha,2i}R_{\alpha,1j}R_{\alpha,2k}+R_{\alpha,2i}R_{\alpha,2j}R_{\alpha,1k}\right)\right.\\
          +\kappa v_S \left(R_{\alpha,2i}R_{\alpha,1j}R_{\alpha,1k}+R_{\alpha,1i}R_{\alpha,2j}R_{\alpha,1k}+R_{\alpha,1i}R_{\alpha,1j}R_{\alpha,2k}\right)\\
              \left.+6\lambda_H v \left(R_{\alpha,1i}R_{\alpha,1j}R_{\alpha,1k}\right)+6\lambda_Sv_S\left(R_{\alpha,2i}R_{\alpha,2j}R_{\alpha,2k}\right)\right]\,.
      \end{aligned}
  \end{align}
  \end{subequations}
The quartic couplings yield
\begin{subequations}
\begin{align}
  &\begin{aligned}
      \includegraphics[width = 0.2\textwidth]{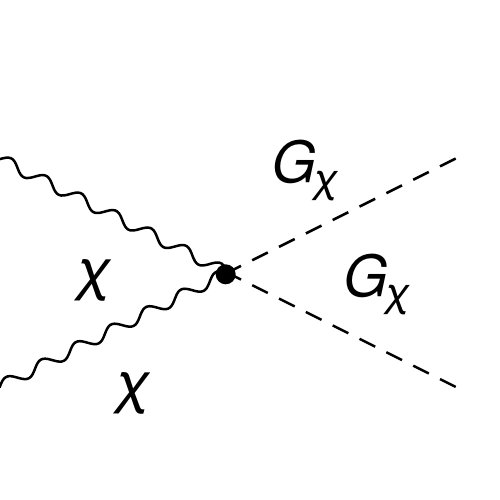}
  \end{aligned}
  =\ii 2 g_{\chi}^2 g_{\mu\nu}\,,\quad
  &\begin{aligned}
      \includegraphics[width = 0.2\textwidth]{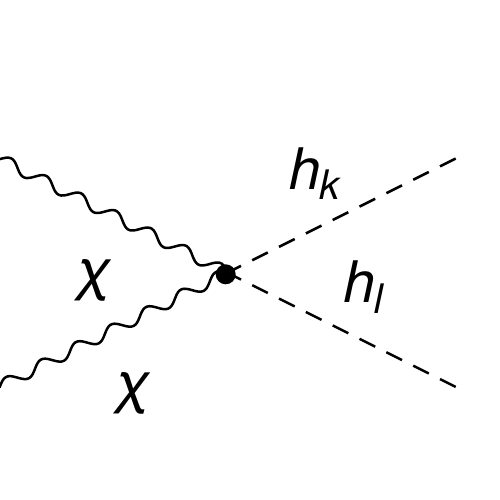}
  \end{aligned}
  =\ii 2 g_{\chi}^2 g_{\mu\nu} R_{\alpha,k2}R_{\alpha,l2}\,,\quad\\
  &\begin{aligned}
      \includegraphics[width = 0.2\textwidth]{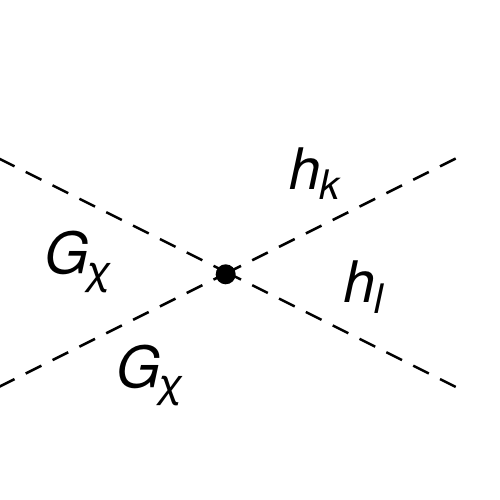}
  \end{aligned}
  =
  \begin{aligned}      
      \ii \frac{g g_{\chi}c_{\alpha}s_{\alpha}}{2 m_W m_{\chi}}R_{\alpha,k1}R_{\alpha,l1}\left(m_{h_2}^2-m_{h_1}^2\right)\\
      -\ii\frac{g_{\chi}^2}{m_{\chi}^2}R_{\alpha,k2}R_{\alpha,l2}\left(m_{h_2}^2 c^2_{\alpha} + m_{h_1}^2s^2_{\alpha}\right)\,.
  \end{aligned}       
\end{align}
\end{subequations}
%%%%%%%%%%%%%%%%%%%%%%%%%%%%%%%%%%%%%%%%%%%%%%%%%%%%%%%%%%%%
%\vspace*{1cm}
%\bibliographystyle{h-physrev}
%\bibliography{direct.bib}

%%%%%%%%%%%%%%%%%%%%%%%%%%%%%%%%%%%%%%%%%%%%%%%%%%%%%%%%%%

\end{document}